\documentclass[a4paper,fleqn,usenatbib]{mnras}

\usepackage{mathptmx}
\usepackage{txfonts}

\usepackage[T1]{fontenc}
\usepackage{ae,aecompl}

\usepackage{graphicx}	
\usepackage{amssymb}	
\usepackage{longtable}
\usepackage{pdflscape}
\usepackage{natbib}
\usepackage{textcomp}
\usepackage{ulem}
\newcommand{\teff}{$T_\textrm{eff}$}
\newcommand{\logg}{$\log g$}

\newcommand{\luminosity}{$\log ({L_\star/L_{\odot}})$}
\newcommand{\kms}{km\,s$^{-1}$}
\newcommand{\vsini}{$v$\,sin\,$i$}
\newcommand{\vrad}{$v_{\rm rad}$}
\newcommand{\halpha}{H$\alpha$}
\newcommand{\hbeta}{H$\beta$}
\newcommand{\hgamma}{H$\gamma$}

\newcommand{\hsigma}{H$\sigma$}

\title[Spectroscopy of normal A and Am stars]{HERMES spectroscopy of normal A and Am stars}

 \author[O. Trust et al.]{
Otto Trust$^{1}$\thanks{E-mail: otrust@must.ac.ug},
Edward Jurua$^{1}$\thanks{E-mail: ejurua@must.ac.ug},
Peter De Cat$^{2}$,
Santosh Joshi$^{3}$, 
 and Patricia Lampens$^2$
 \\
 $^{1}$Department of Physics, Mbarara University of Science and Technology, P.O. Box 1410, Mbarara, Uganda\\
$^{2}$Royal Observatory of Belgium, Ringlaan 3, B-1180 Brussel, Belgium\\
$^{3}$Aryabhatta Research Institute of Observational Sciences, Manora Peak, Nainital- 263002, India
 }

\date{Accepted XXX. Received YYY; in original form ZZZ}

\pubyear{2020}

\begin{document}
\label{firstpage}
\pagerange{\pageref{firstpage}--\pageref{lastpage}}
\maketitle

\begin{abstract}
The nominal {\it Kepler} mission provided very high-precision photometric data. Using these data, interesting phenomena such as spots, and ``hump and spike'' features were observed in the light curves of some normal A and metallic lined A stars (Am stars). However, the connection between such phenomena and the chemical peculiarity of the Am stars is still unclear. In order to make progress on these issues, its important to collect high-resolution spectroscopic data to determine their fundamental parameters and individual chemical abundances. In this paper, we present a spectroscopic study of a sample of ``hump and spike'' stars in the nominal {\it Kepler} field. We used data collected with the High Efficiency and Resolution Mercator \'{E}chelle Spectrograph (\textsc{HERMES}). We determined the spectral type of these stars and obtained the atmospheric stellar parameters such as effective temperatures, surface gravities, projected rotational, microturbulent and radial velocities. We also performed a detailed individual chemical abundance analysis for each target. We confirmed KIC\,3459226 and KIC\,6266219 as Am stars, KIC\,9349245 as a marginal Am star, while KIC\,4567097, KIC\,4818496, KIC\,5524045, KIC\,5650229, KIC\,7667560, and KIC\,9272082 are non-Am stars. To estimate their evolutionary phases, all the stars were placed in the Hertzsprung-Russell (HR) diagram. Based on their spectral classification and chemical abundance pattern, we reclassified KIC\,6266219 (previously treated as chemically normal) as an Am star (kA3hA7mF1) and KIC\,9272082 (previously treated as Am) as non-Am.
\end{abstract}



%
\section{Introduction}

Over the previous few decades, many space missions were lauched, such as Microvariability and Oscillations of Stars \citep[MOST;][]{RUCINSKI2003371}, Convection, Rotation and planetary Transits \citep[CoRoT;][]{2006ESASP.624E..34B}, {\it Kepler} \citep{borucki10} and Transiting Exoplanet Survey Satellite \citep[TESS;][]{1.JATIS.1.1.014003}, which are mainly searching for exo-planets and {enabling} asteroseismology of pulsating stars.
More projects like PLAnetary Transits and Oscillations of stars \citep[PLATO;][]{2014ExA....38..249R} and Wide Field Infrared Survey Telescope \citep[WFIRST;][]{Gehrels_2015} will be launched in the near future. To understand the characteristics of exo-planets, the stellar parameters of the host stars are relevant \citep{2013Natur.503..381H}. For pulsating stars, especially those with solar-like oscillations, asteroseismology has played a big role in determining the stellar parameters using high-precision photometric data from space telescopes. Even when high-precision photometric data are available, ground-based spectroscopy is still essential in determining complementary stellar parameters of most stars and it is still widely used \citep[e.g;][]{2005A&A...437.1127S, 2015A&A...576A..94S, 2018A&A...619L..10G, 2019OAst...28...68K, 10.1093/mnras/staa138}. For the intermediate mass main-sequence (MS) stars (A- and F-type), where chemically peculiar (CP) stars are common (about 10\%), spectroscopy plays a big role in determining the individual chemical abundances in addition to the atmospheric stellar parameters \citep[e.g;][]{2008A&A...483..891F, 2008A&A...483..567G, 2008A&A...479..189G, 2010A&A...523A..71G}.
 
The CP stars exhibit atmospheric abundances which are  significantly different from the solar values. Based on the magnetic field strength and strength of absorption lines in their optical spectra, \citet{1974ARA&A..12..257P} categorised the CP  stars into four major groups: CP1 stars (the metallic-line or Am/Fm stars), CP2 stars (the magnetic Ap stars), CP3 stars (the Mercury-Manganese or HgMn stars) and CP4 stars (the He-weak stars). The CP1 (Am) are the most numerous CP stars followed by CP2 (Ap) stars in the intermediate mass MS part of the Hertzsprung-Russell (HR) diagram. The Am stars are characterized by under-abundance of certain light elements (such as Ca and Sc) and excess of metals, in addition to the very weak or non-detectable magnetic fields in general \citep{1970PASP...82..781C, 1974ARA&A..12..257P, 2007AstBu..62...62R}. This implies that for Am stars, three spectral classifications are always given: (i) based on the Balmer lines, which give good estimates of effective temperature (\teff), (ii) based on the Ca\,\textsc{ii}\,K-line ($\lambda\lambda$ 393.4\,nm)
which gives an earlier spectral type since it is weaker than for normal stars and (iii) based on {the} metallic lines which give a later spectral type because they are enhanced relative to normal stars. Am stars whose spectral types are derived from the Ca\,\textsc{ii}\,K-line and metallic lines differ by 5 or more subtypes are called classical Am stars. For marginal Am stars, this difference is less than 5 subtypes. The CP2 stars are known for large excesses (up to several orders of magnitude) of heavy elements such as Si, Hg or the rare-earth elements Sr, Cr, Eu, Nd, Pr etc \citep{1970PASP...82..781C, 1974ARA&A..12..257P, 2000BaltA...9..253K, 2007AstBu..62...62R}. The CP2 stars are also characterized by the presence of well organized magnetic fields with strengths of up to several tens of kG \citep{2007A&A...475.1053A}, which could be of fossil origin \citep{2004Natur.431..819B}. The CP3 stars possess enhanced Hg\,\textsc{ii} ($\lambda\lambda$ 398.4\,nm) and/or Mn\,\textsc{ii} lines and weak lines of light elements (such as He, Al and N) in their spectra. The majority of CP3 stars are very slow rotators with \vsini\,$\leq$\,20\,\kms\ and are also characterised by weak or non-detectable magnetic fields \citep{1974ARA&A..12..257P, 1997A&A...319..928S, 1998ApJ...504..533B}. The CP4 stars are characterised by anomalously weak He\,\textsc{i} lines in their spectra. They are also slow rotators with magnetic field strengths of upto the orders of 1\,kG \citep{1974PASP...86...67J, 1974ARA&A..12..257P}.
 
The chemical peculiarities in CP stars are thought to mainly originate from the interplay between radiative levitation and gravitational settling, a process known as atomic diffusion \citep{1970ApJ...160..641M,1970ApJ...162L..45W,khokh, hui,2000ApJ...529..338R, 2003ASPC..305..199T, 2011sf2a.conf..253T}. Based on this theory, in the absence of mixing, light elements drift under the influence of gravity and are seen as under-abundant while heavy elements are radiatively driven outward (reflected as overabundant). This theory requires calm and stable atmospheres which are facilitated by the slow rotation of CP stars \citep{2008JKAS...41...83T,2008A&A...483..891F, stateva,Abt2009} and by strong magnetic fields in the case of Ap stars. In addition to turbulent mixing, another actor in the process could be weak mass loss \citep{2000ApJ...529..338R, 2001ApJ...558..377R,2004ApJ...606..452M,2006ApJ...645..634T}. It has also been reported to reduce the main surface chemical abundance anomalies, in models including atomic diffusion, to the observed levels in CP stars \citep{1983ApJ...269..239M,1986ApJ...311..326M,1996A&A...310..872A,2010A&A...521A..62V, 2011A&A...526A..37V}.  The strong magnetic fields are thought to stabilise the convective material, while slow rotation minimises meridional circulation and in-turn minimises mixing that would counteract atomic diffusion. Rotation plays a big role in constraining the physics of CP stars such as pulsations  \citep{1988AcA....38...61D,1998A&A...334..911S}, overshooting \citep{2004ApJ...601..512B, 2019MNRAS.485.4641C} and the observed chemical abundances \citep{2002ASPC..259..258T, 2014PhDT.......131M}. 

\begin{figure}
  \includegraphics[width=\columnwidth]{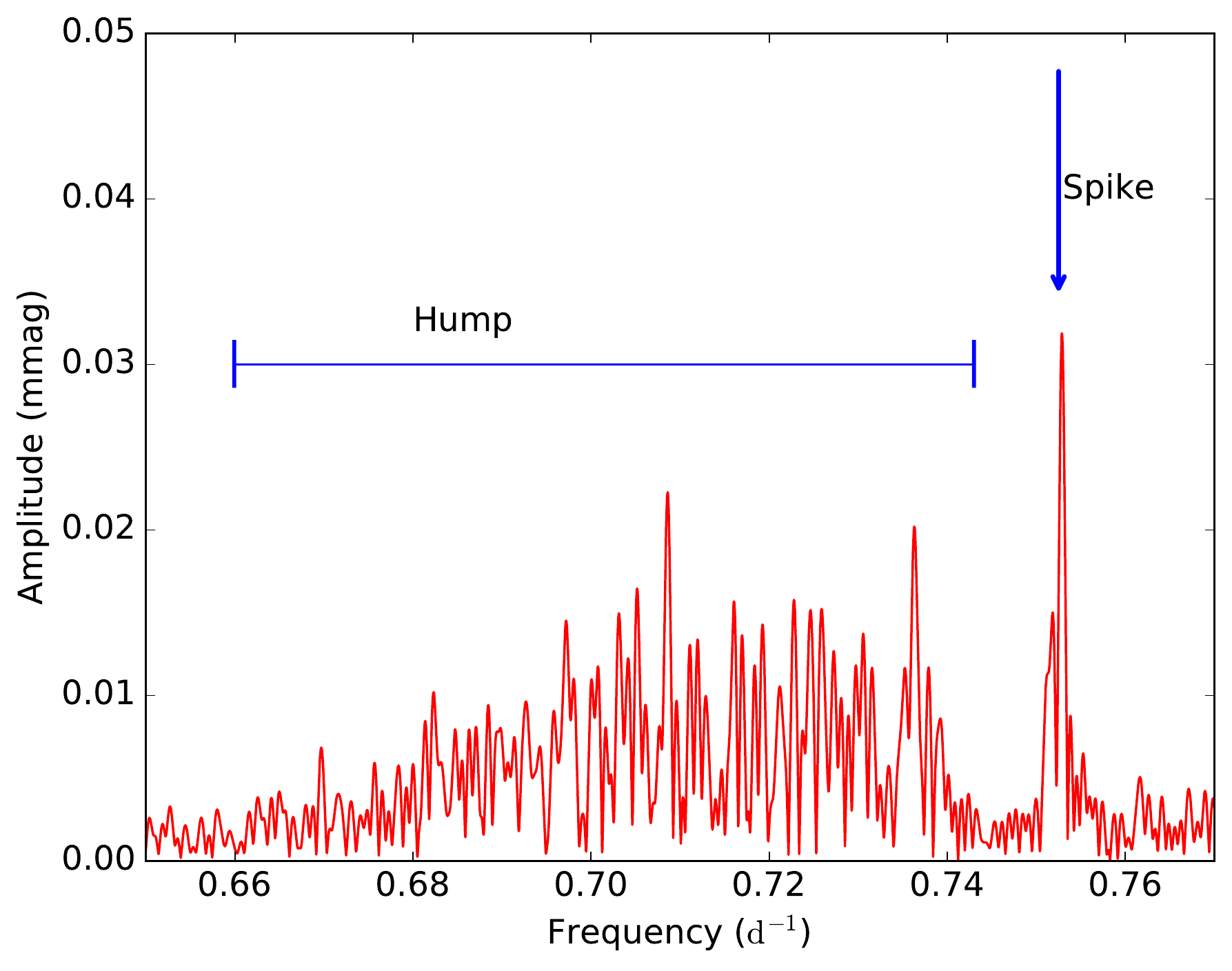}
  \caption{The ``hump and spike'' features in an amplitude spectrum for KIC\,6192566 as reported by \citet{2020MNRAS.492.3143T}.}
  \label{spike}
 \end{figure}

Given the high-precision and almost continuous photometric data from space missions like the nominal {\it Kepler} mission, K2 \citep{2014PASP..126..398H} and TESS, the rotation of A-type (CP and normal) stars has been studied assuming spot induced rotational modulation \citep{1947PASP...59..261K, 2009A&A...506..245M, balona13} and based on the ``hump and spike'' features in their frequency spectra \citep{2018MNRAS.474.2774S, 2020MNRAS.492.3143T}. Briefly, the ``hump and spike'' stars were named by \citet{2018MNRAS.474.2774S}. Their frequency spectra possess a sharp peak (``spike'') on the high frequency side of a broad hump of very close frequencies (``hump'') as shown in Fig.\,\ref{spike} by \citet{2020MNRAS.492.3143T}. The broad humps are induced by Rossby waves ($r$\,modes) and the spikes are the rotation frequencies induced by one or more spots \citep{2018MNRAS.474.2774S}. The ``hump and spike'' stars are a test bed to study the association of the $r$\,modes with rotation and other processes (such as atomic diffusion) that affect their chemical peculiarity. The period of stellar rotation can also be determined using seismic splitting \citep{1981ApJ...244..299S,1981Natur.293..443C} and chromospheric activity, especially Ca\,\textsc{ii} emission \citep{1984ApJ...279..763N}.  However, we obtain a complete rotational profile when the projected rotational velocity (\vsini) is known too. The \vsini\ can be spectroscopically retrieved from the broadening of the absorption lines \citep{1989ssis.book.....K}.

 To search and study the pulsational variabilities in Ap and Am/Fm stars, a dedicated ground-based project, the Nainital-Cape survey, was initiated between astronomers of India and South Africa. With time, astronomers from other countries joined this programme making it a multi-national collaborative project and a number of results have been published \citep[e.g;][]{2000BASI...28..251A, 2001AA...371.1048M,2003MNRAS.344..431J,2006A&A...455..303J, 2009A&A...507.1763J, 2010MNRAS.401.1299J, 2012MNRAS.424.2002J, 2016A&A...590A.116J, 2017MNRAS.467..633J}. The success of the Nainital-Cape survey and the discovery of ``hump and spike'' features in the frequency spectra \citep{balona13,balona15,2018MNRAS.474.2774S} of a number of normal A and Am stars  motivated us to initiate a collaboration with astronomers from India, Uganda and Belgium, to study the phenomena responsible for the observed chemical peculiarities in Am stars. The first results are presented in \citet{2020MNRAS.492.3143T}. This paper is the second of the series whose objective is to characterize the ``hump and spike'' stars. This involves the spectral classification, determination of atmospheric stellar parameters and individual chemical abundances of our targets. 

This paper is organised as follows.  The data are discussed in Sect.\,\ref{data}. Spectral classification is presented in Sect.\,\ref{class}, while atmospheric parameters and individual chemical abundance analyses are given in Sect.\,\ref{atm_para}. The comments on individual stars and their location in the HR diagram are discussed in Sects.\,\ref{comm} and \ref{hr}, respectively, while the conclusion is given in Sect.\,\ref{concl}.

\section{The Data}
\label{data}

\subsection{Selection criteria}
\label{sel}

The majority of stars in this study are part of the samples analysed by \citet{2020MNRAS.492.3143T}. These are normal A and Am stars with ``hump and spike'' features in their frequency spectra  which were selected from \citet{balona13}, \citet{balona15}, and \citet{2016AJ....151...13G}. Of the 171 ``hump and spike'' stars studied by \citet{2020MNRAS.492.3143T}, we selected those with the $\rm V\,\leq 10.5\,mag$ and not reported in the literature as members of binary systems. Stars with $\rm V\,\leq 10.5\,mag$ are bright enough to obtain high-quality HERMES spectra with SNR of at least 100 per pixel in a maximum exposure time of one\,hour. In this paper, we present a spectroscopic analysis for 9 such stars.

\subsection{Observations and Data Reduction}
\label{obs}

The high-resolution spectroscopic observations were made on the nights of 02, 06 and 07 November 2018, using the \textsc{HERMES} \citep{2011A&A...526A..69R} spectrograph mounted at the Cassegrain focus of the \mbox{1.2-m} Mercator telescope located at La Palma, Spain. In a single exposure, this spectrograph records optical spectra in the wavelength ($\lambda$) range 377 to 900\,nm\ across the 55 spectral orders. The resolving power of this instrument in the high-resolution mode is 85\,000 with radial velocity stability of about 50\,m\,s$^{-1}$, and excellent throughput \citep{2011A&A...526A..69R}. The log of the observations collected with HERMES is presented in Table\,\ref{table:log}.

The spectra were reduced using a dedicated HERMES pipeline. The usual reduction procedure for \'{e}chelle spectra was applied. The steps include subtraction of the bias and stray light, flat-field correction, order-by-order extraction, wavelength {calibration with} Thorium-Argon lamps, cosmic rays removal and merging of the orders \citep{2011A&A...526A..69R}. The procedure also provides the SNR, which in our case, falls in the range 30\,--\,110 (at $\lambda=500,\,650$ and 810\,nm) on average and are listed in Table\,\ref{table:log}. Normalisation of all the spectra to the local continuum was performed manually using  the \textit{continuum} task of the \textsc{iraf} package\footnote{http://iraf.noao.edu/}. Finally, all the spectra were systematically corrected for barycentric motion.

For some of the stars, we collected more than one spectrum (Table\,\ref{table:log}). For these stars, all the spectra were median-combined and we performed investigations on the average spectrum. The SNR of the median spectra lies in the range of 110\,--\,220 on average.

\begin{table*}
\caption{The observation log for the target stars.  The right ascension ($\alpha_{\rm J2000}$) and declination ($\delta_{\rm J2000}$) given by \citet{2018yCat.1345....0G} and the V-band magnitude estimated by \citet{2000A&A...355L..27H} are included.}
\label{table:log}
\fontsize{7.5}{9.5}\selectfont
\begin{tabular}{llcccccccr}
\hline\hline\\
KIC&Other&$\alpha_{\rm J2000}$&$\delta_{\rm J2000}$ & V & BJD (2\,450\,000+)&Number& SNR&  Total Exposure\\
No.&Name&(hh mm ss)&(dd mm ss)&(mag)&(day)&of&(@ 500, 650 &Duration\\
&&&&&&Spectra& and 810 nm)&(min)\\
\noalign{\smallskip}
\hline
\noalign{\smallskip}
  3459226 & BD\,$+38^{\circ}3705$ &19 40 50.1201&+38 32 27.0912 &10.11  & 8\,430  &5& 27\,--\,51 & 75.0\\
  4567097 & HD\,184469 &19 32 55.9360& +39 39 45.7237& 7.75  & 8\,429   & 1& 90\,--\,111 & 10.0\\
  4818496 & HD\,177592 &19 03 39.2582&+39 54 39.2370 & 8.07 & 8\,429  & 1& 99\,--\,125 & 12.5\\
  5524045 & BD\,$+40^{\circ}3639$ &19 15 14.7706 & +40 43 45.3009& 9.36 & 8\,429  & 2& 42\,--\,71 & 40.0\\
  5650229 & HD\,226697 &19 56 36.7584 & +40 48 32.0652&  9.92 & 8\,430 &3 & 37\,--\,55 & 42.5\\
  6266219 & TYC3127-2016-1 &18 56 58.4055&+41 36 45.3267& 9.87 & 8\,425 &3& 25\,--\,44 & 60.0\\
  7667560 & HD\,177458 &19 02 49.8015 & +43 21 21.1749& 9.63  & 8\,425 &3 & 23\,--\,51 & 60.0\\
  9272082 & HD\,179458 & 19 10 33.7965 & +45 45 03.9964 & 8.95 & 8\,429  &4 & 55\,--\,86 & 50.0\\
  9349245 & HD\,185658 & 19 38 05.1898 & +45 53 03.8221 & 8.11 & 8\,430 &1 & 70\,--\,104 & 10.0\\
\hline
\end{tabular}
\end{table*}

\section{Spectral classification}
\label{class}

Prior to the determination of atmospheric parameters and chemical abundances, we performed a spectral classification analysis of the targets on the MK classification system \citep{1943assw.book.....M,2009ssc..book.....G}. Spectral classification provides essential details about a star's chemical peculiarity and initial atmospheric parameters \citep{2009ssc..book.....G}. The spectral type and luminosity class are based on the similarity between the observed spectra and those of well-known standards, taking into account essential hydrogen and metal lines. We classified the stars using two data types, (i) spectral energy distribution (SED) and (ii) HERMES spectra.

Using the \textsc{vosa} package \citep{2008A&A...492..277B} and least-squares minimisation technique, we cross-matched the SED (discussed in Sect.\,\ref{SEDs}) of the targets with the template libraries built by \cite{2017ApJS..230...16K} from SDSS spectra. The resulting spectral types are indicated with ``1a'' in Table\,\ref{table:class}.
The uncertainty in the spectral type is about 3 subtypes. 

Classification of the HERMES spectra was performed using the code \textsc{Mkclass} \citep{2014AJ....147...80G}. Given that the observation time is always limited and not all the standards are always visible at the time of observation of the targets, we used the available standard libraries by \citet{2003AJ....126.2048G}. For the observations of these standard libraries, they used the Dark Sky Observatory (DSO) 0.8-m telescope in Northwestern part of North Carolina in combination with the Gray/Miller classification spectrograph having two gratings with 600 and 1200 grooves\,mm$^{-1}$, respectively. The standards  were observed in the violet-green spectral region at a resolution range of 0.15\,--\,0.36\,nm / 2 pixels \citep{2014AJ....147...80G}. 

The spectra for the standards and our sample stars are not from the same spectrograph/grating combination. To approximate the specifications of the standards as close as possible, we truncated the wavelength range, re-binned and convolved the HERMES spectra with a Gaussian of appropriate full width at half maximum (FWHM) determined by the intended final resolution of 0.36\,nm / 2 pixels. 

The metric-distance technique \citep{1994ASPC...60..312L}, which relies on a weighted least-square comparison of the spectra of the program star with those of the MK standard stars, was used \citep{2014AJ....147...80G}.
For each star, the spectral type was determined three times, each time based on different lines, that is; hydrogen lines (\hgamma\ and \hsigma\ lines), metal lines, and the Ca\,\textsc{ii}\,K-line. For a normal star, all the three regions correspond to the same spectral type. However, in the case of the chemically peculiar stars, such as Am stars, the three regions give different spectral types \citep{2009ssc..book.....G}. The obtained spectral types and luminosity classes are indicated with ``1b'' in Table\,\ref{table:class}. The spectral types and luminosity classes of the sample lie in the range B9 to F1 and III to V, respectively.

\begin{table}
\caption{The spectral classification of the target stars.}
\label{table:class}
\begin{tabular}{llcc}
\hline
\hline\\
KIC& Sp type&Comment&References\\
   && &\\
\hline
\noalign{\smallskip}
  3459226 & F2 Dwarf             &       & 1a\\
          & kA2hF0mF3            &       & 1b\\
          & kA2hF0mF2            &       & 2 \\
\noalign{\smallskip}
  4567097 & A0 Giant             &       & 1a\\
          & B9 III               & vgood & 1b\\
          & B9                   &       & 3 \\
\noalign{\smallskip}
  4818496 & A1 Dwarf             &       & 1a\\
          & A1 V                 &  good & 1b\\ 
          & A0                   &       & 3 \\
\noalign{\smallskip}
  5524045 & A1 Dwarf             &       & 1a\\
          & A6 mA0 V metal-weak  &  good & 1b\\ 
          & A3 mA0 IV metal-weak &       & 2 \\
          & A1V                  &       & 4 \\
          & A0.5Va+              &       & 5 \\
\noalign{\smallskip}
  5650229 & A1 Dwarf             &       & 1a\\
          & A8 mA0 V metal-weak  &       & 1b\\
          & A3 mA1 IV metal-weak &       & 2 \\
          & A0 IV-V              &       & 2 \\
          & A2                   &       & 3 \\
          & B9 III or IV         &       & 4 \\
\noalign{\smallskip}
  6266219 & F1 Dwarf             &       & 1a\\
          & kA3hA7mF1            &       & 1b\\
\noalign{\smallskip}
  7667560 & A9 Dwarf             &       & 1a\\
          & A6 mA0 V metal-weak  &  good & 1b\\
          & A                    &       & 3 \\
\noalign{\smallskip}
  9272082 & A9 Dwarf             &       & 1a\\
          & A4 V                 &  good & 1b\\
          & A5 IV-V              &       & 2 \\
          & A0                   &       & 3 \\
          & A3 V                 &       & 4 \\
          & A7                   &       & 6 \\
          & Am                   &       & 7 \\
          & A5m                  &       & 8 \\
\noalign{\smallskip}
  9349245 & F0 Dwarf             &       & 1a\\
          & A8 IV  (Sr)          &  good & 1b\\
          & A8 V (Sr)            &       & 2 \\
          & A5                   &       & 3 \\
          & A7 III               &       & 4 \\
          & Am                   &       & 9 \\
          & F2 III               &       & 10 \\
\hline\end{tabular}

References: (1) This study based on: (a) fitting SED with \cite{2017ApJS..230...16K} template libraries using the \textsc{vosa} tool, and (b) HERMES spectra using \textsc{Mkclass} \citep{2014AJ....147...80G}; (2) \citet{2016AJ....151...13G}; (3) \citet{1993yCat.3135....0C}; (4) \citet{2016A&A...594A..39F}; (5) ``a+'' denotes higher luminosity main-sequence star \citep{2015MNRAS.450.2764N}; (6) \citet{1970A&AS....1....1F}; (7) \citet{1960JO.....43..129B}; (8) \citet{1952ApJ...116..592M}; (9) \citet{catanzaro2015}; (10) \citet{2017MNRAS.470.2870N} 
\end{table}

\section{Atmospheric parameters and chemical abundances}
\label{atm_para}

Photometry and spectroscopy are the traditional methods used to determine atmospheric parameters. In the case of well-studied pulsating stars, asteroseismology has proved to give good results, especially since high-precision photometric data from space missions like CoRoT and {\it Kepler} became available. That does not alter the fact that spectroscopy has always remained robust and relevant in providing atmospheric parameters. However, a good initial guess is very essential to attain convergence during the process of atmospheric parameter determination based on spectral synthesis. 
Some of the methods from which the initial guesses can be obtained include photometric colour indices (Sect.\,\ref{photo}) and SED fitting (Sect.\,\ref{SEDs}).

\subsection{Atmospheric parameters from photometry}
\label{photo}

We computed stellar parameters using the available photometric indices from photometric databases, namely, {\it ubvy$\beta$} \citep{1998A&AS..129..431H}, 2MASS \citep{2003yCat.2246....0C, 2006AJ....131.1163S} and UBV \citep{1985A&AS...59..461O}. All values were retrieved from the General Catalog of Photometric data{\footnote{ 
https://gcpd.physics.muni.cz}}\citep{1997A&AS..124..349M}. The reddening parameter $E{\rm (B-V)}$ was computed from 3D models \citep{bayestar, 2019arXiv190502734G} using {the} GAIA parallaxes \citep{2018yCat.1345....0G} and the stellar galactic coordinates from the SIMBAD database{\footnote{https://simbad.u-strasbg.fr/simbad/}} \citep{2000A&AS..143....9W}. The resulting values of $E{\rm (B-V)}$ are reported in column\,3 of Table\,\ref{uvby}.

In the General Catalog of Photometric data, the Str\"{o}mgren\,$\beta$ indices are available for KIC\,9272082 and KIC\,9349245 only. We calculated \teff\ and \logg\ from {\it ubvy$\beta$} indices using a routine originally written by \cite{moon1985stellar}, based on \cite{1985MNRAS.217..305M} and corrected by \cite{1993A&A...268..653N}. We determined the overall metallicity ([M/H]) using the Str\"{o}mgren $\delta m_o$ index for A-type (A0\,--\,F0) stars \citep{1993A&A...274..391S}. We considered errors (systematic and statistical) of 200\,K, 0.10\,{dex} \citep{1993A&A...268..653N} and 0.13\,dex \citep{1993A&A...274..391S} in \teff, \logg\, and [M/H], respectively. The results based on Str\"{o}mgren photometry are given in columns 4 to 6 of Table\,\ref{uvby}.

Finally, we computed \teff\ from the 2MASS and UBV photometric systems. This is based on the strong correlation of  \teff\ with the (V\,--\,K)$_o$ \citep{2006A&A...450..735M} and (B\,--\,V)$_o$ indices \citep{2000AJ....120.1072S}. Since \teff\ is weakly correlated with \logg\ and [M/H], we fixed $\log g=4.0$\,{dex} and $\rm{[M/H]}=0.0$\,dex (solar metallicity). For each system, we estimated the uncertainties in \teff\ based on the errors of the photometric indices, i.e. $E{\rm (B-V)}$ (0.02\,mag), [M/H] (0.10\,dex) and \logg\ (0.10\,{dex}) \citep{2016MNRAS.458.2307K}. The photometric parameters and their uncertainties are listed in columns 7 and 8 of Table\,\ref{uvby}. The \teff\ (2MASS) and \teff\ (UBV) for KIC\,4567097 and KIC\,6266219 do not agree within $1\sigma$. In addition, \teff\ ({\it ubvy$\beta$}) values for KIC\,9272082 and KIC\,9349245 seem to be about 500\,K higher compared to the values of \teff\ (2MASS) and \teff\ (UBV). The difference in \teff\ could be introduced by the assumption that (V\,--\,K) and (B\,--\,V) are completely \logg\ and metallicity independent \citep{2005ApJ...626..465R}. Some of the target stars in the sample are chemically peculiar and/or variable, which may introduce additional errors in the calibration process \citep{2016A&A...585A..67H}.

\subsection{Atmospheric Parameters from SED}
\label{SEDs}

We can measure stellar parameters from the SED of a star. SEDs are built using spectrophotometry obtained in different wavelengths, ideally from ultraviolet to infrared. For this purpose, we adopted the \textsc{vosa} tool.

\begin{figure}
\centering
\includegraphics[width=0.89\columnwidth]{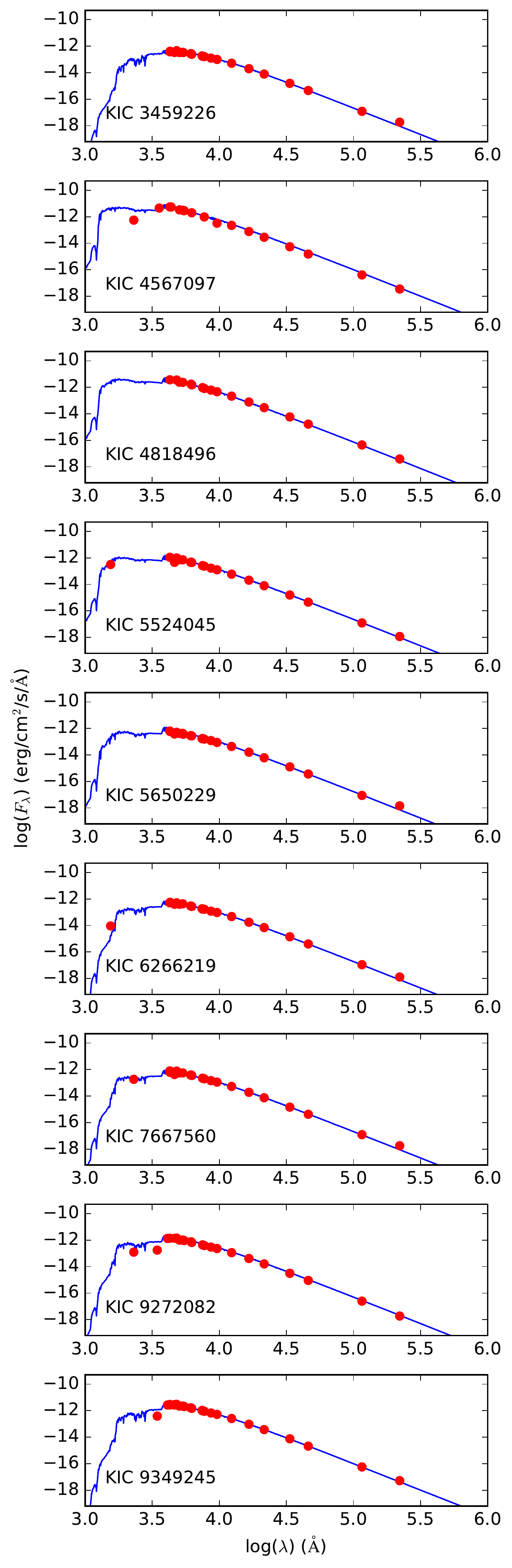}
\caption{The red filled circles represent the SEDs of the target stars. The blue solid lines represent the spectra fitted to the data using the \textsc{vosa} tool.
}
 \label{sed_fig}
\end{figure}

In the first step, the tool automatically searches for spectrophotometric observations in different databases, i.e. the Tycho\,II \citep{2000A&A...357..367H}, 2MASS \citep{2006AJ....131.1163S}, WISE \citep{2010AJ....140.1868W}, and GAIA DR2 \citep{2018A&A...619A.180M} catalogs, complemented with Johnson \citep{1953ApJ...117..313J, 1985A&AS...59..461O}, Str\"{o}mgren \citep{1998A&AS..129..431H}, Galaxy Evolution Explorer \citep[GALEX;][]{2000MmSAI..71.1123B}, Sloan \citep{Brown} and Pan-STARRS1 \citep{2016arXiv161205560C} photometry, when available. We estimated the extinction ($A_{\rm v}$) for each target using $ A_{\rm V} = R_{\rm V} \times E{\rm (B-V)}=3.1\times E{\rm (B-V)}$ \citep{Brown}.  We determined $E{\rm (B-V)}$ as discussed in Sect.\,\ref{photo}. Also the stellar distance is required to construct SEDs. We used distances calculated from the GAIA parallaxes \citep{2018yCat.1345....0G} that are listed in column\,2 of Table\,\ref{uvby}. The SEDs for all the stars under study are shown in Fig.\,\ref{sed_fig} as filled circles. Using ATLAS9 Kurucz ODFNEW/NOVER models \citep{1997A&A...318..841C} in the \textsc{vosa} tool, we performed a least-square fit to these SEDs to obtain \teff, \logg\ and [M/H]. The results are represented by the blue lines in Fig.\,\ref{sed_fig}. The uncertainties calculated by the \textsc{vosa} tool on these parameters are underestimated by a factor of about 2 because they only account for the internal errors on the SED-fitting procedure. This is attributed to the intrinsic limitations of the method \citep{2008A&A...492..277B,2012MNRAS.427..343M, catanzaro2015}. Therefore, to have more realistic error limits, we doubled the uncertainties of the quantities \citep[cf.][]{catanzaro2015}, making the $T_{\rm eff, SED}$ uncertainties comparable to those from other photometric techniques. The results of the SED fitting procedure are listed in columns 9 to 12 of Table\,\ref{uvby}. 

\begin{table*}
\caption{List of basic and photometric parameters. The parallax ($\pi$) determined by \citet{2018yCat.1345....0G}, reddening parameter ($E({\rm B-V})$) computed using 3D dustmaps by \citet{2019arXiv190502734G}, the effective temperature (\teff), surface gravity (\logg) and metallicity ([M/H]) estimated using {\it uvby$\beta$} indices, \teff\ computed from 2MASS, \teff\ derived from UBV, and the estimated \teff, \logg\ and [M/H] from the SED fitting are given.}
\label{uvby}
\fontsize{7.2}{12}\selectfont
\begin{tabular}{lcccccccccc}
\hline
\hline
\noalign{\smallskip}
KIC&$\pi$&$E(B-V)$&\teff&\logg&[M/H]&\teff&\teff&\teff&\logg&[M/H]\\ 
&&& ${}^{uvby\beta}$& ${}^{uvby\beta}$& ${}^{uvby\beta}$& $\rm{}^{2MASS}$&$\rm{}^{UBV}$&$\rm{}^{SED}$&$\rm{}^{SED}$&$\rm{}^{SED}$ \\
No.&(mas)&(mag) &($\pm$\,200\,K)&($\pm$\,0.10\, dex)&($\pm$\,0.13\,dex) &(K)&(K)&($\pm$\,250\,K)&($\pm$\,0.5\,{ dex})&(dex)\\
\noalign{\smallskip}
\hline
\noalign{\smallskip}
  3459226 & 3.3107\,$\pm$\,0.0804 & 0.0303 &-&- &-& 7390\,$\pm$\,280 & 6990\,$\pm$\,250 & 7500& 4.0 & 0.0\,$\pm$\,0.35 \\
  4567097 & 2.0990\,$\pm$\,0.0480 & 0.0581 &- &- &-& 9490\,$\pm$\,510 & 10430\,$\pm$\,350 & 9750 & 3.5 & 0.2\,$\pm$\,0.25 \\
  4818496 & 4.0709\,$\pm$\,0.0366 & 0.0593 &- &- &-& 9580\,$\pm$\,290 &  9400\,$\pm$\,270 & 9500&4.5 & 0.0\,$\pm$\,0.35 \\
  5524045 & 2.1854\,$\pm$\,0.0338 & 0.0449 &- &- &- & 8900\,$\pm$\,360 & 9170\,$\pm$\,260 & 9250&4.5 & 0.0\,$\pm$\,0.35 \\
  5650229 & 1.6616\,$\pm$\,0.0286 & 0.1377 &- &- &-& 9250\,$\pm$\,340 &  9450\,$\pm$\,330 & 9000&3.5 & 0.5\,$\pm$\,0.30 \\
  6266219 & 2.5992\,$\pm$\,0.0362 & 0.0310 &- &- &-& 7930\,$\pm$\,260 &  7160\,$\pm$\,260 & 7750 &3.5 & 0.2\,$\pm$\,0.25 \\
  7667560 & 2.8792\,$\pm$\,0.0347 & 0.0245 &- &- &-& 8310\,$\pm$\,290 & 8340\,$\pm$\,220 & 8000& 3.5 & 0.2\,$\pm$\,0.25 \\
  9272082 & 2.5738\,$\pm$\,0.0386 & 0.0254 & 8470 & 4.13& 0.17& 8010\,$\pm$\,170 &  7920\,$\pm$\,130 & 8000& 3.5 & 0.5\,$\pm$\,0.30 \\
  9349245 & 6.9544\,$\pm$\,0.0343 & 0.0100 & 8130 & 4.27& 0.22& 7780\,$\pm$\,170 &  7690\,$\pm$\,120 & 7500 &3.5 & 0.5\,$\pm$\,0.30 \\
\hline\end{tabular}
\end{table*}

\subsection{Atmospheric Parameters from Spectroscopy}

In this section we evaluate the fundamental astrophysical quantities of our targets, i.e. \teff, \logg, \vsini, microturbulent velocity ($\xi$), and radial velocity (\vrad). 

The first step is to determine \vrad\ by computing the cross correlation function (CCF) using different pre-selected masks built from line lists using the code \textsc{iSpec} \citep{2014A&A...569A.111B, 2019MNRAS.486.2075B}. We computed \vsini\ by comparing the observed spectra with a grid of synthetic spectra, pre-computed with plane-parallel ATLAS9\footnote{http://www.stsci.edu/hst/observatory/crds/castelli\_kurucz\_atlas.html} model atmospheres \citep{2003IAUS..210P.A20C} for different values of \teff\ and of \logg, in the wavelength range 516\,--\,519\,nm (Mg\,\textsc{i} triplet). This comparison was performed with a least-square method based on the \textsc{minuit} minimization package used by the \textsc{girfit} package \citep{2006A&A...451.1053F}. 

For stars with $T_{\rm eff}\geq 8000$\,K, the Balmer lines become more sensitive to \logg\ than to \teff. For such targets (KIC\,4567097, KIC\,4818496, KIC\,5524045, KIC\,5650229, KIC\,6266219, KIC\,7667560, and KIC\,9272082), we used \textsc{girfit} to derive the temperature from the spectral region with a high number of metal lines (495\,--\,570\,nm). The SNR of this spectral region was first improved using the least-square deconvolution (LSD) method described in \cite{2013A&A...560A..37T}. Using \textsc{girfit}, we fitted 10\,--\,15\,nm segments of the denoised spectrum and calculated the average \teff\ for all the segments. For KIC\,3459226 and KIC\,9349245 ($T_{\rm eff}<8000$\,K), we determined \teff\ using hydrogen line profiles. The errors are a contribution of uncertainties encountered in continuum normalization and errors from the fitting procedure. We adopted an error of 100\,K as a contribution from continuum normalization \citep{catanzaro2015}. 

The values of \logg\ were estimated by two different methods depending on \teff\ of each object. For target stars with $T_{\rm eff}>8000$\,K, we computed \logg\ from fitting hydrogen line wings. At  $T_{\rm eff}<8000$\,K the Balmer lines loose their sensitivity to \logg, and so metal lines (Fe\,\textsc{i}/Fe\,\textsc{ii} and/or Mg\,\textsc{i} triplet) were used instead. 

\begin{figure}
\centering
 \includegraphics[width=\columnwidth]{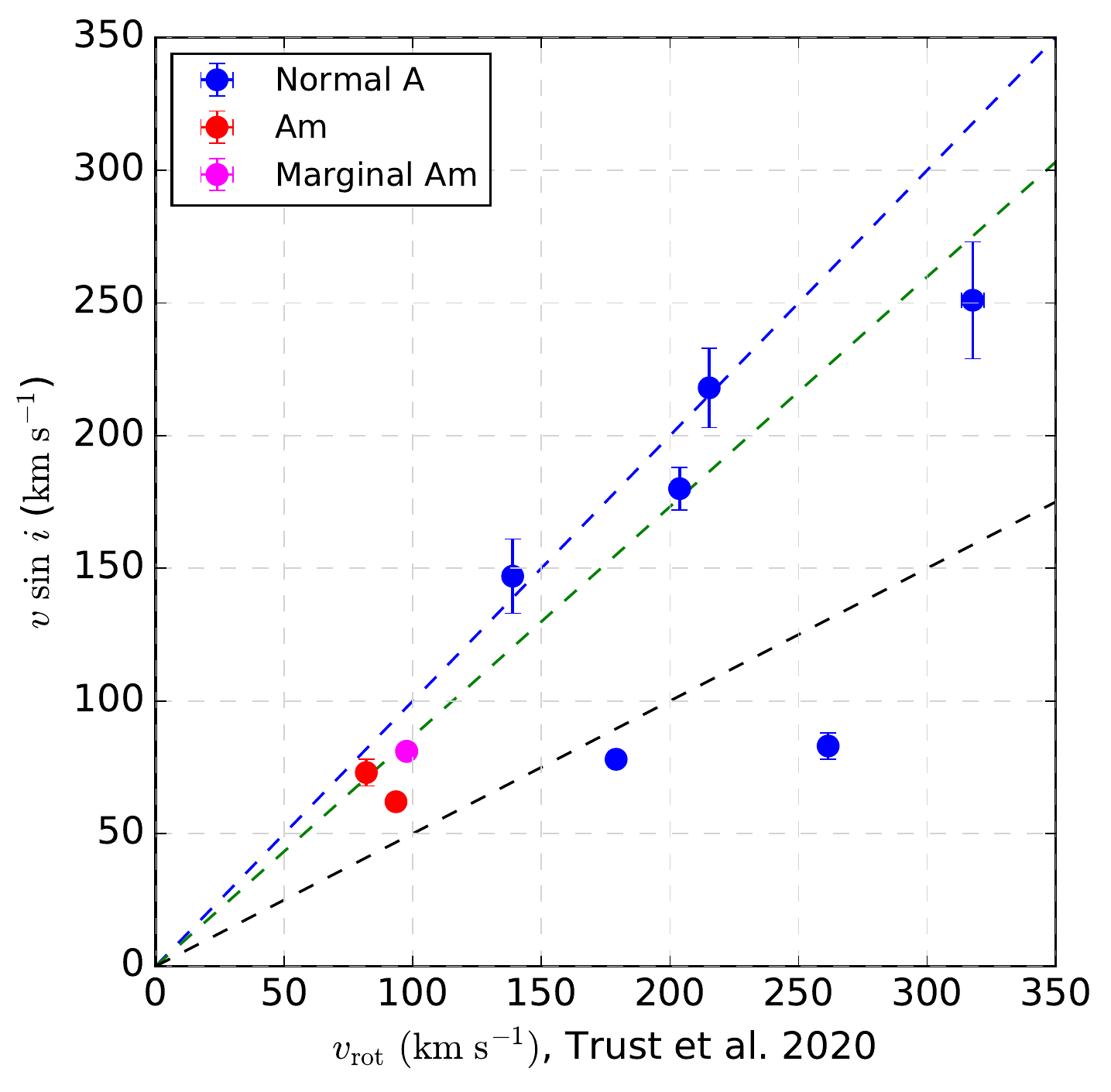}
 \caption{ The projected rotational velocity \vsini\ as computed in this study compared with the equatorial rotational velocity $v_{\rm rot}$ by \citet{2020MNRAS.492.3143T}. The blue, red, and magenta dots are normal A, Am, and marginal Am stars, respectively. The blue dashed line represents sin\,$i=1$ while the green and black ones correspond to $i =$\,60\textdegree, and 30\textdegree, respectively.}
 \label{vrot_fig}
\end{figure}

\begin{figure*}
\centering
 \includegraphics[width=\textwidth]{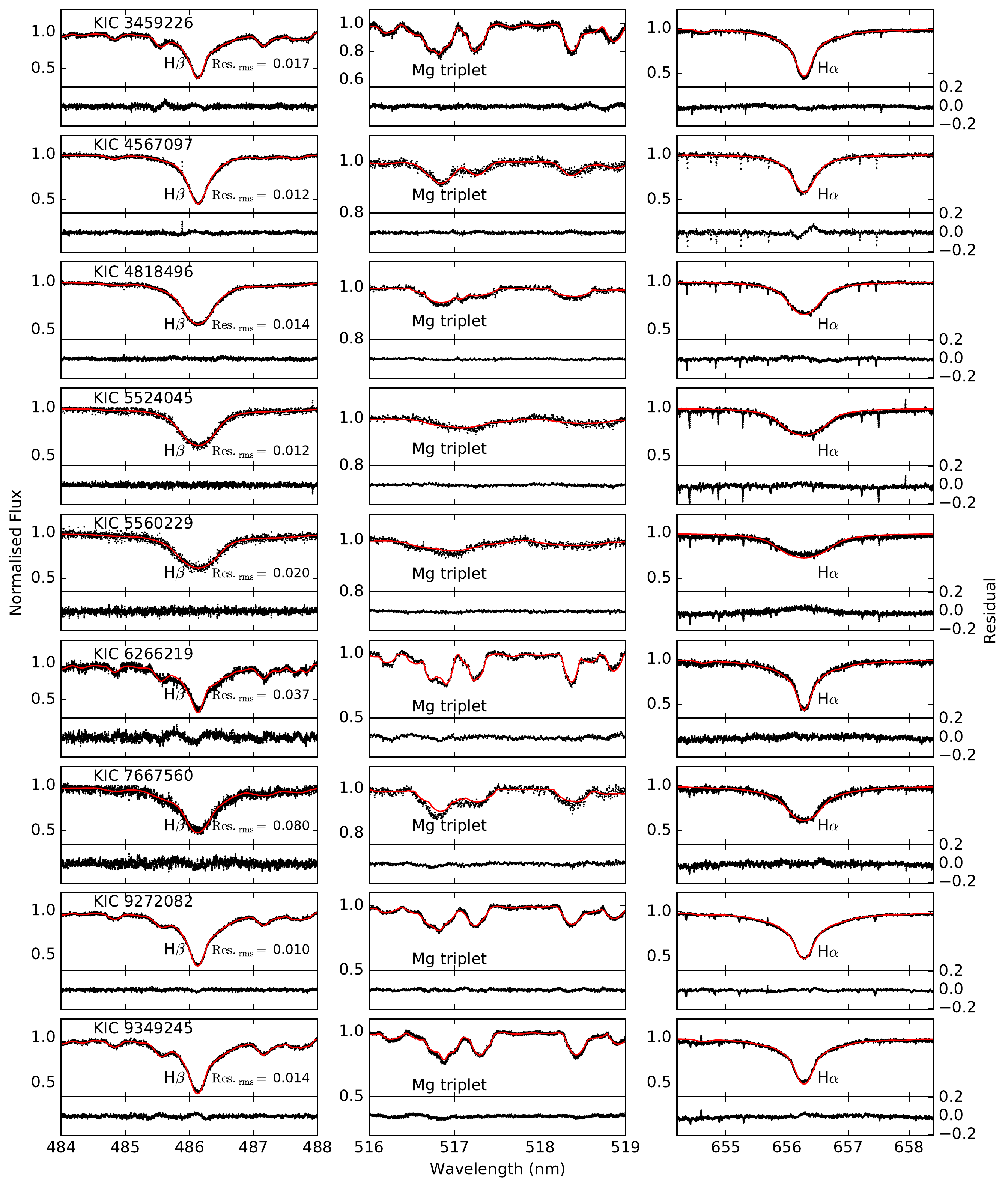}
 \caption{The \hbeta\ (left), Mg\,\textsc{i} triplet (middle) and \halpha\ (right) spectral regions for the target stars. The observed (black) and synthetic spectra (red) are shown. The root mean square value of the residuals ($\rm Res._{rms}$) for each spectrum is given in the bottom-right corner of each \hbeta\ panel. }
 \label{hb_fig}
\end{figure*}

\begin{figure}
\centering
 \includegraphics[width=\columnwidth]{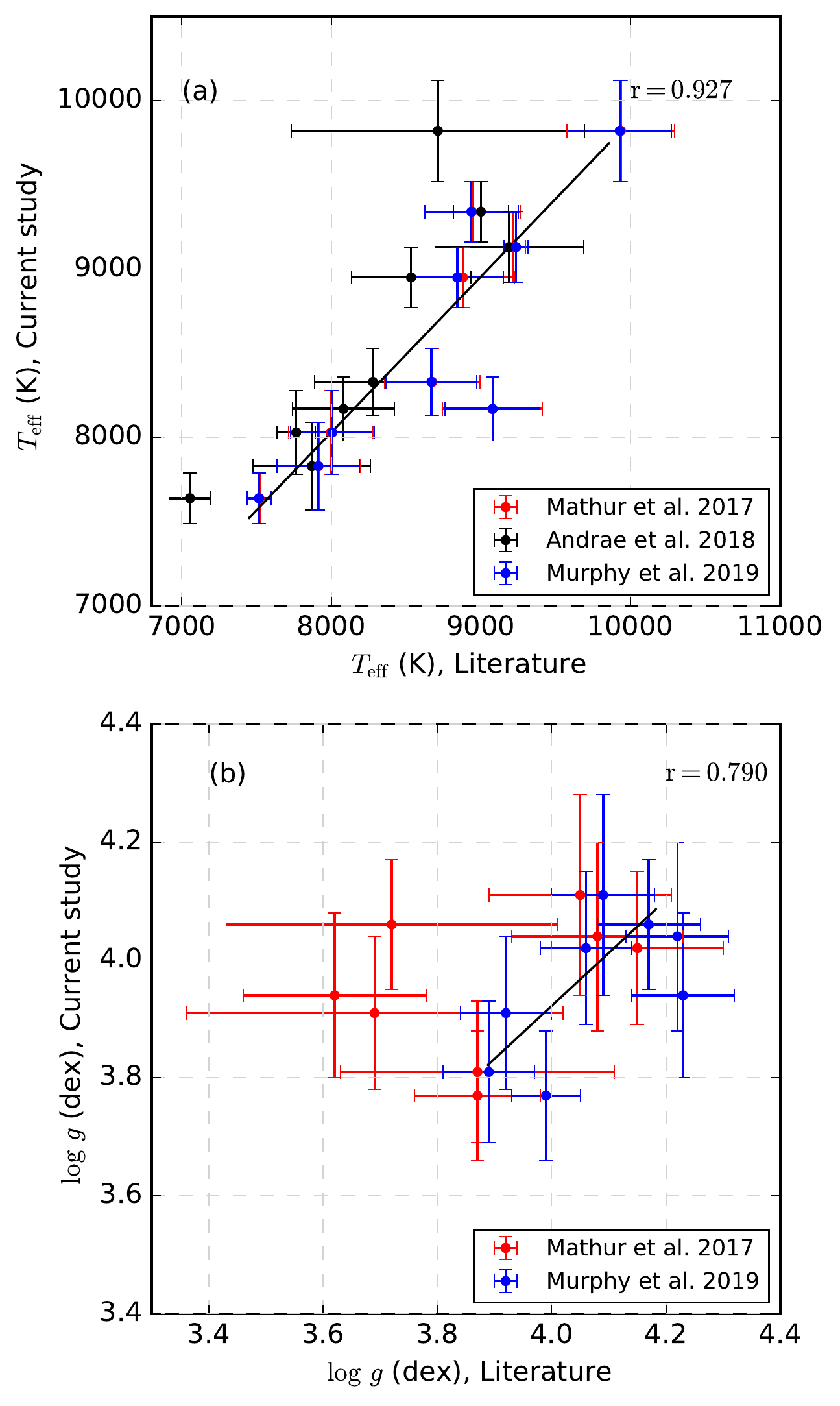}
 \caption{The \teff\ and \logg\ from the current study compared with those reported by \citet{2017ApJS..229...30M} (red), \citet{2018A&A...616A...8A} (black), and  \citet{10.1093/mnras/stz590} (blue). The respective correlation coefficient, r, with the weighted average of the literature values is given in the top-right corner of each panel.
 }
 \label{fig:comp_T}
\end{figure}

We uniquely obtained $\xi$ by fitting iron lines using the MOOG radiative transfer code \citep{2012ascl.soft02009S}, ATLAS9 model atmospheres, the Vienna Atomic Line Database (VALD) line-list \citep{1999A&AS..138..119K} and \citet{2009ARA&A..47..481A} solar abundances, all incorporated in an integrated software \textsc{iSpec}. We determined the initial guess for $\xi$ from the relation \citep{gebran14}, 
\begin{equation}
 \xi=3.31 \times \exp\left[-\left(\log\left(\frac{T_{\rm eff}}{8071.03}\right)^2/0.01045\right)\right].
\end{equation}
Fig.\,\ref{hb_fig} shows the \hbeta\ (left panels), Mg\,\textsc{i} triplet (middle panels), and \halpha\ (right panels) spectral regions for the target stars (black) and the synthetic spectra (red) computed with the final results of the atmospheric parameters. The results for the atmospheric fundamental quantities and their errors are listed in Table\,\ref{table:spec}.  The spectroscopic \teff\ values differ from those estimated using photometric indices ({\it ubvy$\beta$}, 2MASS and UBV) and SED by an average temperature of about 360\,K, 180\,K, 350\,K and 180\,K, respectively. The spectroscopic \logg\ and [M/H] values are more consistent with those estimated from {\it ubvy$\beta$} photometry  (average $\Delta$\logg\,$_{{uvby}\beta}=0.18$\,dex and $\Delta$[M/H]\,$_{{uvby}\beta}=0.10$\,dex) than those from SED  (average $\Delta$\logg\,$_{\rm SED}=0.5$\,dex and $\Delta$[M/H]\,$_{\rm SED}=0.23$\,dex). Our final results (from spectroscopy) agree in general within the error limits with those from previous studies. The literature values for which this is not the case are  shown in italics in Table\,\ref{table:spec}. Particularly, we compared our \teff\ and \logg\ with those computed by \citet{2017ApJS..229...30M, 2018A&A...616A...8A, 10.1093/mnras/stz590} as shown, in panels (a) and (b) of Fig.\,\ref{fig:comp_T}, respectively. Based on the correlation coefficient (presented in the top right corner of each panel) with the weighted average of the literature values, \teff\ is more consistent than \logg. The \vsini\ values are excellently consistent with the equatorial rotational velocity ($v_{\rm rot}$) computed by \citet{2020MNRAS.492.3143T} as shown in Fig.\,\ref{vrot_fig}, where all the stars are positioned either on or below the sin\,$i=1$ line.

\begin{table*}
\centering
\caption{List of stellar parameters from spectroscopic analysis. The different columns show: identification number, effective temperatures (\teff), surface gravity (\logg), metallicity ([M/H]), microturbulent velocity ($\xi$), projected rotational velocity (\vsini) and radial velocity (\vrad). The literature values that do not agree within 1$\sigma$ with the results from this study are given in italics.
}
\label{table:spec}
\begin{tabular}{lccccccr}
\hline
\hline
\noalign{\smallskip}
KIC&\teff&\logg&[M/H]&$\xi$&\vsini&\vrad&References\\ 
\noalign{\smallskip}
No.&(K)&(dex) &(dex) &(\kms)&(\kms) &(\kms)&\\
\noalign{\smallskip}
\hline
\noalign{\smallskip}
  3459226
  & 7640\,$\pm$\,150 & 3.94\,$\pm$\,0.14 & 0.32\,$\pm$\,0.08 & 3.78\,$\pm$\,0.18 & 73\,$\pm$\,5&-28.8\,$\pm$\,2.2&1 \\
  &7410\,$\pm$\,139 & 3.79\,$\pm$\,0.14 &&&$<$\,120& -24.1\,$\pm$\,24.6 &2\\
  &7519\,$\pm$\,83  &{\it 3.62\,$\pm$\,0.16}&&&&  &3\\
  &{\it 7055\,$\pm$\,140}&&&&&  &4\\
  &7516\,$\pm$\,80  & 4.23\,$\pm$\,0.09 &&&&  &5\\
  \hline
  \noalign{\smallskip}
  4567097 & 9820\,$\pm$\,300 & 3.90\,$\pm$\,0.12& 0.08\,$\pm$\,0.13& 1.64\,$\pm$\,0.35 & 83\,$\pm$\,5 & 7.1\,$\pm$\,2.4 & 1\\
  &9935\,$\pm$\,360&4.21\,$\pm$\,0.20&&&&&3\\
  &8712\,$\pm$\,980&&&&&  &4\\
  &9928\,$\pm$\,347&&&&&  &5\\
  \hline
  \noalign{\smallskip}
  4818496 & 9340\,$\pm$\,180 & 3.91\,$\pm$\,0.13& 0.25\,$\pm$\,0.10&  2.26\,$\pm$\,0.23 & 180\,$\pm$\,8 &11.8\,$\pm$\,5.4&1\\
  &8941\,$\pm$\,323&3.69\,$\pm$\,0.33&&&&&3\\
  &9000\,$\pm$\,185&&&&&&4\\
  &8936\,$\pm$\,312&3.92\,$\pm$\,0.08&&&&&5\\
  &&&&&& 12.8\,$\pm$\,0.8 &6\\
  \hline
  \noalign{\smallskip}
  5524045 & 9130\,$\pm$\,210 & 3.77\,$\pm$\,0.10& -0.15\,$\pm$\,0.09& 2.16\,$\pm$\,0.17 & 218\,$\pm$\,15 & 3.6\,$\pm$\,13.0 & 1\\
  &9267\,$\pm$\,169 &3.87\,$\pm$\,0.11&&&201\,$\pm$\,58 & 7.8\,$\pm$\,36.2 &2\\
  &9217\,$\pm$\,82&3.87\,$\pm$\,0.16&&&&&3\\
  &9189\,$\pm$\,497&&&&&&4\\
  &9235\,$\pm$\,80 &{\it 3.99\,$\pm$\,0.06}&&&&&5\\
  &{\it 9500\,$\pm$\,200}&4.0\,$\pm$\,0.20&&{\it 1.0\,$\pm$\,0.4}&215\,$\pm$\,7 &  &7\\
  \hline
  \noalign{\smallskip}
  5650229 & 8950\,$\pm$\,180 & 3.81\,$\pm$\,0.12 & 0.19\,$\pm$\,0.10 & 2.02\,$\pm$\,0.19 & 251\,$\pm$\,22 & -9.2\,$\pm$\,1.2 & 1\\
  & 9898\,$\pm$\,420 &3.88\,$\pm$\,0.12&&&243\,$\pm$\,33 & -9.3\,$\pm$\,22.1 &2\\ 
  &8879\,$\pm$\,345&3.87\,$\pm$\,0.24&&&&&3\\
  &8533\,$\pm$\,400&&&&&&4\\
  &8842\,$\pm$\,309&3.89\,$\pm$\,0.08&&&&  &5\\
  \hline
  \noalign{\smallskip}  
  6266219 & 8030\,$\pm$\,250 & 4.11\,$\pm$\,0.17 &  0.63\,$\pm$\,0.11 & 3.88\,$\pm$\,0.23 & 62\,$\pm$\,3 & 2.0\,$\pm$\,1.5 & 1\\
  &7996\,$\pm$\,282&4.05\,$\pm$\,0.16&&&&  &3\\
  &7765\,$\pm$\,130&&&&&&4\\
  &8007\,$\pm$\,280&4.09\,$\pm$\,0.09&&&&&5\\
  \hline
  \noalign{\smallskip}
  7667560 & 8330\,$\pm$\,200 & 4.04\,$\pm$\,0.16 & 0.16\,$\pm$\,0.09 & 2.41\,$\pm$\,0.29 & 147\,$\pm$\,14 &-20.2\,$\pm$\,4.8 &1\\
  &8675\,$\pm$\,320& 4.08\,$\pm$\,0.15&&&&  &3\\
  &8278\,$\pm$\,390&&&&&  &4\\
  &8669\,$\pm$\,303&4.22\,$\pm$\,0.09&&&&  &5\\
  \hline
  \noalign{\smallskip}
  9272082 & 8170\,$\pm$\,190 & 4.02\,$\pm$\,0.13 & 0.17\,$\pm$\,0.10 & 3.63\,$\pm$\,0.20 & 78\,$\pm$\,3 & 5.7\,$\pm$\,1.7 &1\\
  &{\it 7551\,$\pm$\,262}&3.91\,$\pm$\,0.12&&&$<$\,120& -20.8\,$\pm$\,18.3 &2\\
  &{\it 9077\,$\pm$\,334}& 4.15\,$\pm$\,0.15&&&& &3\\
  &8081\,$\pm$\,340&&&&&&4\\
  &{\it 9078\,$\pm$\,317}&4.06\,$\pm$\,0.08&&&&&5\\
  \hline
  \noalign{\smallskip}
  9349245 & 7830\,$\pm$\,260 & 4.06\,$\pm$\,0.11 & 0.37\,$\pm$\,0.12 & 3.33\,$\pm$\,0.14 & 81\,$\pm$\,3 &-36.2\,$\pm$\,6.3 &1\\
  &7683\,$\pm$\,79 &3.87\,$\pm$\,0.11&&&$<$\,120& -15.0\,$\pm$\,23.6 &2\\
  &7914\,$\pm$\,279& 3.72\,$\pm$\,0.29&&&& &3\\
  &7870\,$\pm$\,393&&&&&&4\\
  &7913\,$\pm$\,276&4.17\,$\pm$\,0.09&&&&&5\\
  &{\it 8300\,$\pm$\,200}&4.00\,$\pm$\,0.30&&3.1\,$\pm$\,0.5&80\,$\pm$\,3&&8\\
  &{\it 8000\,$\pm$\,200}&{\it 3.7\,$\pm$\,0.1} &&2.9\,$\pm$\,0.1&82\,$\pm$\,2&&9\\
\hline\end{tabular}\\
References: (1) This study; (2) \citet{2016A&A...594A..39F}; (3) \citet{2017ApJS..229...30M}; (4) \citet{2018A&A...616A...8A}; (5) \citet{10.1093/mnras/stz590}; (6) \citet{2006AstL...32..759G}; (7) \citet{2015MNRAS.450.2764N}; (8) \citet{catanzaro2015};  (9) \citet{2017MNRAS.470.2870N} 
\end{table*}

\subsection{Individual chemical abundances}
\label{abu}
The individual chemical abundances were determined based on direct fitting of the theoretical profiles of individual spectral lines. Since some targets are fast rotators and others are CP stars, lines are often blended and direct spectrum synthesis takes into account possible blends \citep{2019MNRAS.488.2343R}. We used the \textsc{SynthV\_NLTE} code \citep{2019ASPC..518..247T} and a grid of atmosphere models pre-computed with the LLmodels code \citep{2004A&A...428..993S}.  This was implemented through an IDL visualization program, \textsc{BinMag6} \citep{2018ascl.soft05015K}. During spectral synthesis, local thermodynamic equillibrium (LTE) was assumed. For oxygen abundance analysis, we excluded the O\,\textsc{i} triplet (around 777.1\,--\,777.5\,nm), since it is affected by non-LTE effects \citep{2000A&A...359.1085P}. For the same reason, we never considered Na\,\textsc{i} lines at 589.0 and 589.6\,nm \citep{2009PASJ...61.1165T} in our analysis. We adopted spectral line lists and atomic parameters from the 3D release of the Vienna Atomic Line Database (VALD3; \citealt{Ryabchikova_2015}). 

We divided each spectrum into several intervals, each $\approx5$\,nm wide, and derived the abundances in each interval by performing a $\chi^2$ minimization of the difference between the observed and synthetic spectrum. The average individual chemical abundances are listed in Table\,\ref{table:chem}. We used the errors on \teff, \logg, \vsini, $\xi$, position of the continuum, and accuracy of the oscillator strengths ($\log({\rm gf})$) of the considered lines to derive abundance uncertainties. For each element, the errors on abundance associated with the errors on each of the \teff, \logg, \vsini, $\xi$, position of the continuum, and the $\log({\rm gf})$ were added together in quadrature to get the total uncertainty. We note that the abundances of elements whose lines only appear in less than three segments should be taken with caution since they were computed from few lines.

\begin{landscape}
\begin{table}
\caption{Individual chemical abundances inferred for our target stars. The solar abundances are listed in the last column \citep{2009ARA&A..47..481A}. The values are expressed in the form log$(N_{\rm el}/N_{\rm Tot})$. The number of segments from which each element was determined is given in brackets.}
\label{table:chem}
 \begin{tabular}{lcccccccccc}
\hline
\hline
\noalign{\smallskip}
&KIC\,3459226&KIC\,4567097 & KIC\,4818496 & KIC\,5524045 & KIC\,5650229 & KIC\,6266219 & KIC\,7667560 & KIC\,9272082 & KIC\,9349245 & Solar\\
\noalign{\smallskip}
\hline
\noalign{\smallskip}
  C & -3.93\,$\pm$\,0.16 (3) & -4.08\,$\pm$\,0.11 (3) & -3.63\,$\pm$\,0.10 (4) & -3.77\,$\pm$\,0.09 (6) & -3.87\,$\pm$\,0.10 (6) & -3.63\,$\pm$\, 0.12 (9) & -4.16\,$\pm$\,0.09 (5) & -3.87\,$\pm$\,0.13 (5) & -3.75\,$\pm$\,0.09 (6) & -3.57\\
  O & -3.10\,$\pm$\,0.08* (1) & -3.22\,$\pm$\,0.15* (1) & -2.48\,$\pm$\,0.09* (1) & -2.61\,$\pm$\,0.10* (2) & -3.22\,$\pm$\,0.16* (2) & -3.06\,$\pm$\,0.10* (2) & -3.01\,$\pm$\,0.15* (1) & -2.95\,$\pm$\,0.09* (1)  & -2.60\,$\pm$\,0.10* (1) & -3.31\\
  Na & -4.12\,$\pm$\,0.07* (1) & -4.01\,$\pm$\,0.20* (1) & -5.02\,$\pm$\,0.10* (1) & - & - & -4.75\,$\pm$\,0.11* (2) & -4.92\,$\pm$\,0.13* (2) & -5.01\,$\pm$\,0.15* (2) & -4.89\,$\pm$\,0.09* (2) & -5.76\\
  Mg & -4.23\,$\pm$\,0.20 (6) & -4.30\,$\pm$\,0.14 (6) & -3.95\,$\pm$\,0.11 (7) & -4.47\,$\pm$\,0.10 (8) & -4.19\,$\pm$\,0.22 (4) & -4.08\,$\pm$\,0.11 (9) &  -4.31\,$\pm$\,0.13 (5) & -4.35\,$\pm$\,0.11 (10) & -4.08\,$\pm$\,0.09 (9) & -4.40\\
  Al & -4.61\,$\pm$\,0.09* (1) & - & - & - & -4.98\,$\pm$\,0.11* (1) & - & - & -5.25\,$\pm$\,0.13* (1) & - & -5.55\\
  Si & -3.91\,$\pm$\,0.16 (8) & -4.27\,$\pm$\,0.14* (2) & -4.00\,$\pm$\,0.11* (1) & -4.01\,$\pm$\,0.11 (3) & -3.71\,$\pm$\,0.15 (3) & -3.82\,$\pm$\,0.11 (10) & -3.83\,$\pm$\,0.17* (2) & -4.40\,$\pm$\,0.09 (12) & -3.73\,$\pm$\,0.10 (5) &-4.49\\
  S & -4.22\,$\pm$\,0.10* (1) & - & -  & -  & -  & -4.29\,$\pm$\,0.09 (3) & -  & -4.48\,$\pm$\,0.15* (1) & -  &  -4.88 \\
  K & -5.92\,$\pm$\,0.09* (1) & - & - & - & - & -6.63\,$\pm$\,0.09* (1) & - & -6.97\,$\pm$\,0.09* (1) & - & -6.97\\
  Ca & -6.05\,$\pm$\,0.14 (11) & -5.89\,$\pm$\,0.12 (3) & -5.18\,$\pm$\,0.14 (3) & -5.78\,$\pm$\,0.18 (5) & -5.79\,$\pm$\,0.12 (5) & -5.67\,$\pm$\,0.10 (23) & -5.48\,$\pm$\,0.20 (6) & -5.68\,$\pm$\,0.09 (18) & -5.38\,$\pm$\, 0.10 (20) & -5.66\\
  Sc & -9.35\,$\pm$\,0.12 (4) & -8.92\,$\pm$\,0.17 (5) & -9.04\,$\pm$\,0.12 (4) & -9.06\,$\pm$\,0.10 (4) & -8.82\,$\pm$\,0.24 (4) & -9.11\,$\pm$\,0.16 (7) & -8.66\,$\pm$\,0.13 (5) & -8.70\,$\pm$\,0.10 (10) & -8.96\,$\pm$\,0.09 (3) & -8.85\\
  Ti &  -6.88\,$\pm$\,0.17 (30) & -7.36\,$\pm$\,0.10 (12) & -6.87\,$\pm$\,0.12 (11) & -7.22\,$\pm$\,0.10 (19) & -7.16\,$\pm$\,0.16 (15) & -6.57\,$\pm$\,0.09 (24) & -6.96\,$\pm$\,0.13 (26) & -6.97\,$\pm$\,0.12 (27) &  -6.77\,$\pm$\,0.10 (23) & -7.05\\
  V & -7.80\,$\pm$\,0.10* (2) & -7.87\,$\pm$\,0.13* (1) & -7.85\,$\pm$\,0.15* (1) & - & -7.86\,$\pm$\,0.12* (2) & -6.98\,$\pm$\,0.12 (4) &  -8.22\,$\pm$\,0.14* (2) &  -8.01\,$\pm$\,0.10 (3) & -7.32\,$\pm$\, 0.12 (5) & -8.07\\
  Cr & -6.04\,$\pm$\,0.13 (5) & -6.58\,$\pm$\,0.09 (7) & -6.39\,$\pm$\,0.09 (7) & -6.95\,$\pm$\,0.23 (7) & -6.30\,$\pm$\,0.10 (7) & -5.53\,$\pm$\,0.11 (17) & -6.36\,$\pm$\,0.13 (10) & -6.30\,$\pm$\,0.09 (19) & -6.15\,$\pm$\,0.09 (18) & -6.36\\
  Mn & -6.16\,$\pm$\,0.19 (3) & -6.37\,$\pm$\,0.11* (1) & -6.38\,$\pm$\,0.12* (1) & -6.74\,$\pm$\,0.12 (3) & -  & -6.31\,$\pm$\,0.10 (3) & -6.89\,$\pm$\,0.12 (3) & -7.02\,$\pm$\,0.23 (5) & -6.18\,$\pm$\,0.08 (5) & -6.57\\
  Fe & -3.89\,$\pm$\,0.11 (79) & -4.63\,$\pm$\,0.11 (17) & -4.24\,$\pm$\,0.13 (38) & -4.68\,$\pm$\,0.11 (33) & -4.27\,$\pm$\,0.13 (31) & -3.82\,$\pm$\,0.10 (83) & -4.25\,$\pm$\,0.14 (43) & -4.61\,$\pm$\,0.09 (43) & -4.18\,$\pm$\,0.09 (44) & -4.50\\
  Co & -5.94\,$\pm$\,0.09 (6) & - & -  & - & -6.34\,$\pm$\,0.11* (1) & -5.71\,$\pm$\,0.13 (8) & - & -6.67\,$\pm$\,0.16 (3) & - & -7.01\\
  Ni & -5.10\,$\pm$\,0.10 (16) & -5.46\,$\pm$\, 0.14* (1) & -6.22\,$\pm$\,0.19* (1) & -5.23\,$\pm$\,0.09* (1) & -4.67\,$\pm$\,0.16 (21) & -5.14\,$\pm$\,0.09 (21) & -5.37\,$\pm$\,0.15 (6) & -5.88\,$\pm$\,0.09 (9) & -5.29\,$\pm$\,0.09 (9) &-5.78\\
  Cu &  -7.05\,$\pm$\,0.13 (3) & - & - & - & -  & -6.37\,$\pm$\,0.23 (3) & - &  -7.17\,$\pm$\,0.21 (3) & - & -7.81\\
  Zn & -7.02\,$\pm$\,0.09 (3) & - & - & -  & - & -7.27\,$\pm$\,0.13* (1) & - & -7.29\,$\pm$\,0.09* (1) & -7.17\,$\pm$\,0.09* (1) & -7.44\\
  Sr & -8.12\,$\pm$\,0.11 (4) & -9.36\,$\pm$\,0.12* (2) & - & -9.04\,$\pm$\,0.15* (1) & -8.84\,$\pm$\,0.10* (1) & -8.29\,$\pm$\,0.09* (2) & -8.84\,$\pm$\,0.22* (2) & -8.75\,$\pm$\,0.15* (1) & -8.49\,$\pm$\,0.14 (3) & -9.13\\
  Y & -8.83\,$\pm$\,0.19 (6) &  - & -9.81\,$\pm$\,0.11* (1) & -9.83\,$\pm$\,0.11* (1) & - & -9.03\,$\pm$\,0.19 (6) & -9.78\,$\pm$\,0.09 (3) & -9.62\,$\pm$\,0.11 (6) & -9.42\,$\pm$\,0.09 (5) &-9.79\\
  Zr & -8.51\,$\pm$\,0.21 (4) & - & - & - & -9.51\,$\pm$\,0.21 (3) & -8.98\,$\pm$\,0.10 (3) & -9.09\,$\pm$\,0.11* (1) & -9.18\,$\pm$\,0.14* (2) & -9.12\,$\pm$\,0.17 (13) &-9.47\\
  Ba & -8.69\,$\pm$\,0.19 (5) & - & - & - & -10.06\,$\pm$\,0.11* (2) &  -8.44\,$\pm$\,0.17 (5) & -9.77\,$\pm$\,0.16 (3) & -9.45\,$\pm$\,0.14 (5) &  -8.76\,$\pm$\,0.14 (3) & -9.82\\
  La & -8.61\,$\pm$\,0.13 (8) & - & - & - & - & -8.91\,$\pm$\,0.21 (10) & - & -10.63\,$\pm$\,0.10 (4) & - & -10.90\\
  Ce & -8.93\,$\pm$\,0.18 (7) & - & - & - & - & -8.61\,$\pm$\,0.11 (9) & - & -10.02\,$\pm$\,0.12 (5) & - & -10.42\\
  Pr & -9.79\,$\pm$\,0.10* (1) & - & - & - & - & -9.07\,$\pm$\,0.10* (2) & - & - & - & -11.28\\
  Nd & -8.73\,$\pm$\,0.11 (5) & - & - & - & - & -8.79\,$\pm$\,0.18 (7) & - & - & - & -10.58\\
\hline
\end{tabular}\\
*These elements were only found in less than three segments.
\end{table}
\end{landscape}

\begin{figure}
\centering
 \includegraphics[width=\columnwidth]{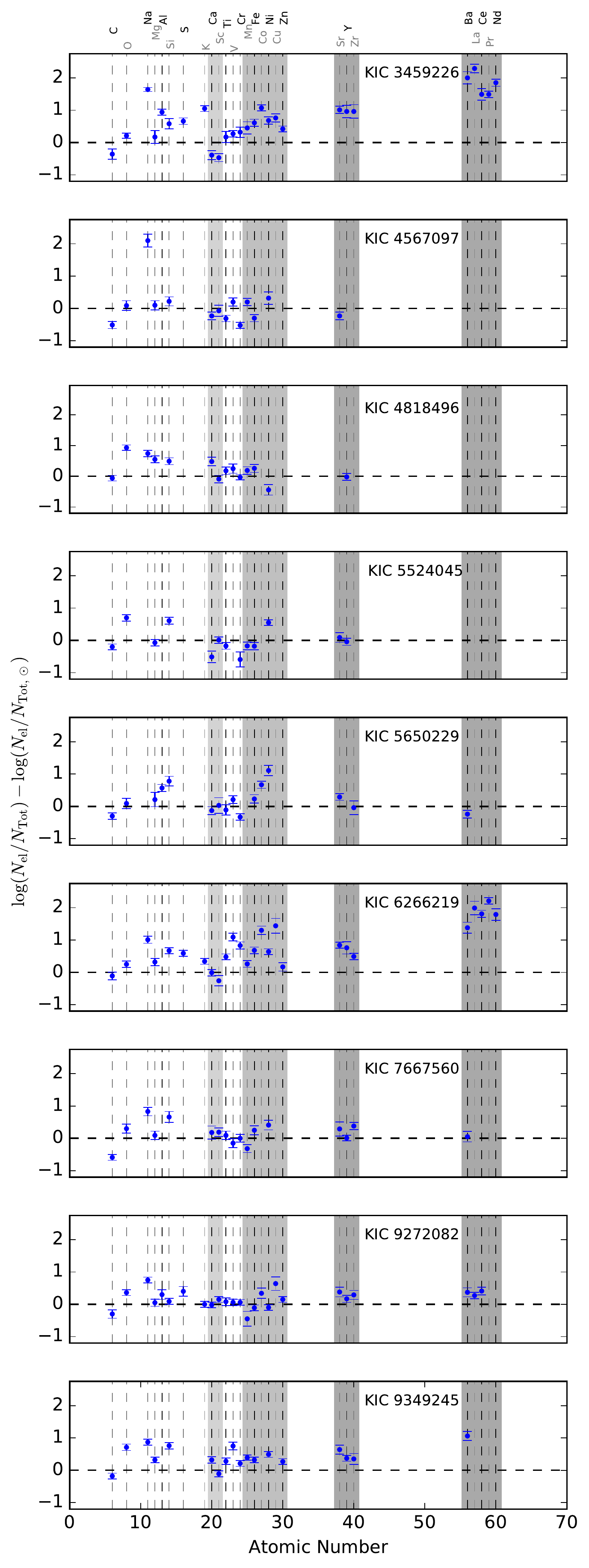}
 \caption{The individual chemical abundance patterns for our targets. The horizontal dashed line represents solar abundance \citep{2009ARA&A..47..481A}. The gray background highlights the light elements (Ca and Sc) and heavy elements (Sr, Y, Zr, Ba, La, Ce, Pr and Nd) important for Am classification. }
 \label{fig:abundances}
\end{figure}

\section{Comments on individual stars}
\label{comm}
In this section we present the findings of the abundance  analysis obtained for each target. The derived abundances and the estimated uncertainties are expressed as log$(N_{\rm el}/N_{\rm Tot})$ and are reported in Table\,\ref{table:chem}. Fig.\ref{fig:abundances} shows the abundance patterns for each star relative to the solar abundances \citep{2009ARA&A..47..481A}.

\subsection{KIC 3459226 (BD +38 3705)}

This star was classified as kA2hF0mF2 by \citet{2016A&A...594A..39F} using LAMOST data  \citep{2012RAA....12.1197C,2012RAA....12..723Z}. We classified this target as kA2hF0mF3 using \textsc{Mkclass} with HERMES data. Calcium and scandium are $\approx$\, 0.4 and $\approx$\,0.5\,dex below solar values, respectively, while the iron-peak and heavy elements such as Sr, Y, Zr, Ba, La, Ce, Pr, and Nd are clearly overabundant (values up to $\approx$\,2\,dex). These over-abundances and spectral classification confirm the Am peculiarity of this star.

\subsection{KIC 4567097 (HD 184469)}

KIC\,4567097 was reported in the Henry Draper (HD) catalog \citep{1993yCat.3135....0C} as a B9 star. Using \textsc{Mkclass}, we obtained a spectral type of B9\,III.  No abundance analysis was found in the literature. The calcium abundance is $\approx$\,0.23\,dex below solar value while scandium and the iron-peak elements are generally solar-like. Only few absorption lines were available and used to determine abundances. Based on the spectral class and chemical abundances obtained, we propose to consider KIC\,4567097 as a non-Am star. 

\subsection{KIC 4818496 (HD 177592)}

This object is listed in the HD catalog as an A0 star \citep{1993yCat.3135....0C} while our analysis classifies it as A1\,V. The \teff\ in this study is slightly higher than the values in the literature while our \logg\ value is consistent (see Table\,\ref{table:spec}). We found this star to be a fast rotator with a \vsini\ value of 180\,$\pm$\,8\,\kms\ which is consistent with the value computed by \citet{2020MNRAS.492.3143T}. O, Na, Mg, Si, and Ca are overabundant (0.5\,--\,0.9\,dex) while Sc and iron-peak elements are, on average, of solar content. Based on these results,
we classify this star as a (chemically) normal A-type star.

\subsection{KIC 5524045 (BD +40 3639)}

KIC\,5524045 was first classified as A0.5\,Va+ (``a+'' denotes higher luminosity MS star) by \citet{2015MNRAS.450.2764N} and subsequently as A1\,V by \citet{2016A&A...594A..39F}. They are both in excellent agreement with our classification as a A1\,V star. We observed this star to be fast rotating with \vsini\,=\,219\,$\pm$\,12\,\kms. Similar rotational velocities, 215\,$\pm$\,7\,\kms\ \citep{2015MNRAS.450.2764N} and 201\,$\pm$\,58\,\kms\ \citep{2016A&A...594A..39F} were published before. All the measured elements are on average of solar type. The top panel of Fig.\,\ref{fig:comp-2015} shows that our abundances are in agreement with those determined by \citet{2015MNRAS.450.2764N}. Based on the abundance pattern, we classify this star as normal.

\subsection{KIC 5650229 (HD 226697)}

This object was first listed as an A2 star in the HD catalog \citep{1993yCat.3135....0C}. Recently, \citet{2016A&A...594A..39F} classified it as a B9\,III or IV star. Our analysis indicates a spectral type as A1\,IV. The most recent study \citep{10.1093/mnras/stz590} presents the star's \teff\ and \logg\ as 8842\,$\pm$\,309\,K and 3.89\,$\pm$\,0.08\,cm\,s$^{-2}$, respectively, which agrees with our result (8920\,$\pm$\,190\,K and 3.82\,$\pm$\,0.11\,cm\,s$^{-2}$). This star is a fast rotator whose \vsini\ was first reported as 243\,$\pm$\,33\,\kms\ \citep{2016A&A...594A..39F}, which is consistent with our outcome of 252\,$\pm$\,24\,\kms.   With exception of Al, Si, Co, and Ni which are overabundant (with an average abundance of $\approx0.8$\,dex), the abundances of the other elements are solar-like and hence we consider this star as non-Am.

\subsection{KIC 6266219 (TYC 3127-2016-1)}

KIC\,6266219 was listed among A-type stars in the nominal {\it Kepler} field \citep{balona13}. We obtained its spectral type as kA3hA7mF1. The available \teff\,=\,7996\,$\pm$\,282\,K and \logg\,=\,4.05\,$\pm$\,0.16\,cm\,s$^{-2}$ \citep{2017ApJS..229...30M}, \teff\,=\,7765\,$\pm$\,130\,K \citep{2018A&A...616A...8A} and \teff\,=\,8007\,$\pm$\,280\,K and \logg\,=\,4.09\,$\pm$\,0.09\,cm\,s$^{-2}$ \citep{10.1093/mnras/stz590} are consistent with our results.  This object is a moderate rotator with \vsini\ of 60\,\kms\ which agrees with the value given by \citet{2020MNRAS.492.3143T}. The chemical abundances of Ca and Sc are solar-like while iron-peak and heavy elements are overabundant up to $\approx$\,2\,dex. Based on the spectral type and chemical abundance pattern, we classify KIC\,6266219 as an Am star.  

\subsection{KIC 7667560 (HD 177458)}

KIC\,7667560 was listed in the HD catalog as an A-type star by \citet{1993yCat.3135....0C}. Our spectral classification analysis revealed a spectral type of A3\,IV. The \teff\ and \logg\ values published by \citet{2017ApJS..229...30M}, \citet{2018A&A...616A...8A}, and \citet{10.1093/mnras/stz590} and are fully consistent with the values obtained in this study. Ca, Sc, iron-peak and heavy elements, on average, have solar-like abundances. Na ($\approx0.83$\,dex) and Si ($\approx0.66$\,dex) are overabundant. The chemical abundance pattern does not show Am characteristics. 

\subsection{KIC 9272082 (HD 179458)}

This star was first classified as Am by \citet{1952ApJ...116..592M} and subsequently by \citet{1960JO.....43..129B}. The most recent classification as A3\,V, by \citet{2016A&A...594A..39F}, matches our result, A4\,V. As shown in Table\,\ref{table:spec}, only the \teff\ computed by \citet{ 2018A&A...616A...8A} using GAIA observations agrees with the \teff\ obtained in this study. Our \logg\ value agrees well with the values: 3.91\,$\pm$\,0.12\,dex \citep{2016A&A...594A..39F}, 4.15\,$\pm$\,0.15\,dex \citep{2017ApJS..229...30M}, and 4.06\,$\pm$\,0.08\,dex \citep{10.1093/mnras/stz590}. With exception of Na ($\approx0.75$\,dex) which is overabundant, abundances of the rest of the elements are solar-like. Based on the spectral classification and the chemical abundance pattern, we consider this star as non-Am.

\subsection{KIC 9349245 (HD 185658)}

\begin{figure}
\centering
 \includegraphics[width=\columnwidth]{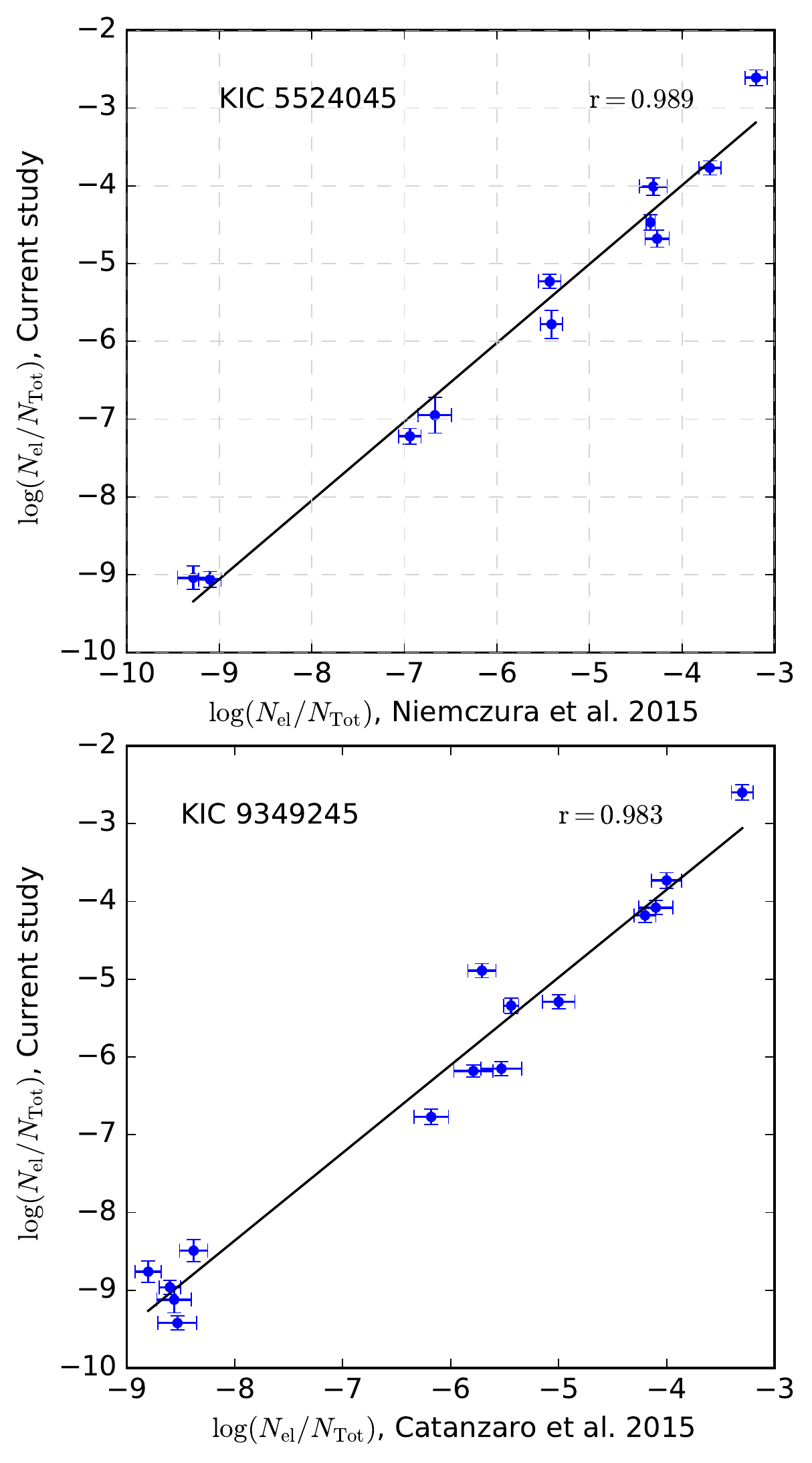}
 \caption{A comparison of our individual chemical abundances for KIC\,5524045 (top panel) with values by \citet{2015MNRAS.450.2764N}
 and  KIC\,9349245 (bottom panel) with those determined by \citet{catanzaro2015}. The corresponding correlation coefficient, r, is given in the top-right corner of each panel.}
 \label{fig:comp-2015}
\end{figure}

KIC\,9349245 is a MS A-type star as reported in various studies (e.g. \citealt{1993yCat.3135....0C}, \citealt{2016AJ....151...13G}, and \citealt{2016A&A...594A..39F}) and it is included in the catalog of chemically peculiar stars \citep{2009A&A...498..961R}. We determined its spectral type as A8 IV (Sr) which is similar to the one given by \citet{2016AJ....151...13G}. The atmospheric parameters and velocities ($\xi$, \vsini\ and \vrad), within error limits, generally agree with values in the literature except for \teff\ of \citet{catanzaro2015} (see Table\,\ref{table:spec}). The Ca and Sc abundances are solar-like, while iron-peak and heavy metals are moderately overabundant (with an average abundance of $\approx0.5$\,dex). The chemical abundance pattern implies a marginal Am star. Similar observations were made by \citet{catanzaro2015}. The bottom panel of Fig.\,\ref{fig:comp-2015} shows a good agreement between the two studies.

\section{Location in the HR Diagram}
\label{hr}

A precise positioning of both Am and non-Am stars in the HR diagram is useful for investigating possible systematic differences in the region occupied by these objects with regard to normal A stars. Moreover, all the targets studied in this paper are "hump and spike" stars and their position is useful to constrain the location of stars with $r$\,modes in the HR diagram. 

\begin{figure}
\centering
 \includegraphics[width=\columnwidth]{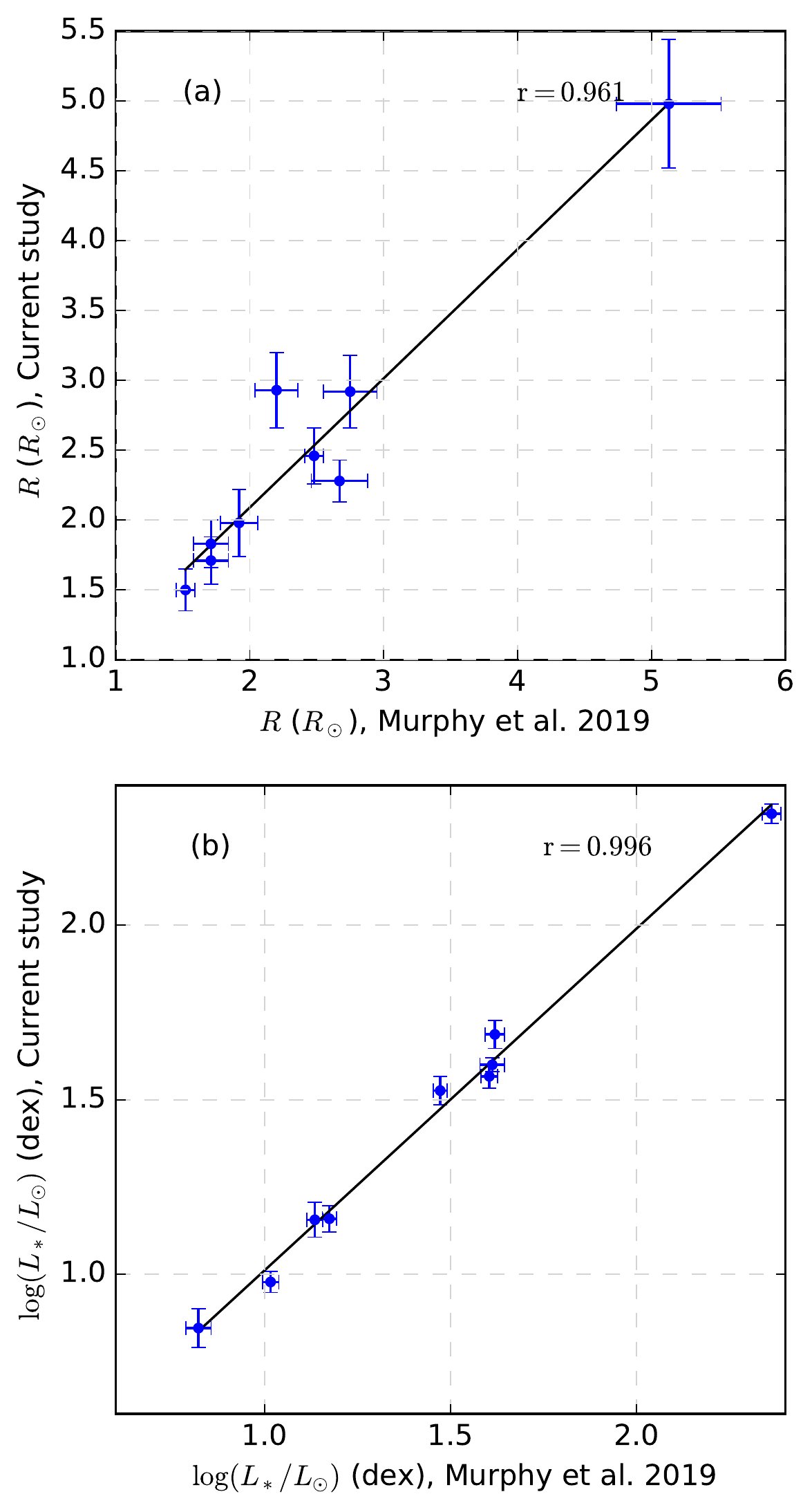}
 \caption{Stellar radius and \luminosity\ from the current study compared with those determined by \citet{10.1093/mnras/stz590}. The respective correlation coefficient, r, is given in the top-right corner of each panel.}
 \label{fig:comp_L}
\end{figure}

The luminosity (\luminosity) of the target stars was calculated using the standard technique. We computed extinction from $ A_{\rm V}=3.1\,\times\,E({\rm B-V})$, where $E({\rm B-V})$ was obtained as discussed in Sect.\,\ref{photo}. The bolometric correction (BC) was calculated using the temperature dependent function of \citet{1996ApJ...469..355F} after a revision by \citet{Torres_2010}, while the absolute magnitude ($M_{\rm v}$) was determined using the GAIA parallaxes \citep{2018yCat.1345....0G}. The uncertainties in \luminosity\ result from uncertainties in V-band indices, parallax and \teff. Having obtained precise \teff\ and \luminosity\ values, we also determined the stellar radius using the standard relation (from the Stefan–Boltzmann law) \citep{1884AnP...258..291B, 2015PhST..165a4027P, Montambaux2018},
\begin{equation}
  {{\rm log}\left(\frac{R}{R_{\odot}}\right)=0.5{\rm log}\left(\frac{L_\star}{L_{\odot}}\right)}-2.0{\rm log}\left(T_{\rm eff}\right)+7.52340473.
\end{equation}
The results and their uncertainties are listed in Table\,\ref{table:lum}. Our values for $R$ (except for KIC\,4818496 and KIC\,9272082) and \luminosity\ are in good agreement with those determined by \citet{10.1093/mnras/stz590}, as shown in panels (a) and (b) of Fig.\,\ref{fig:comp_L}, respectively. 

\begin{figure}
\centering
 \includegraphics[width=\columnwidth]{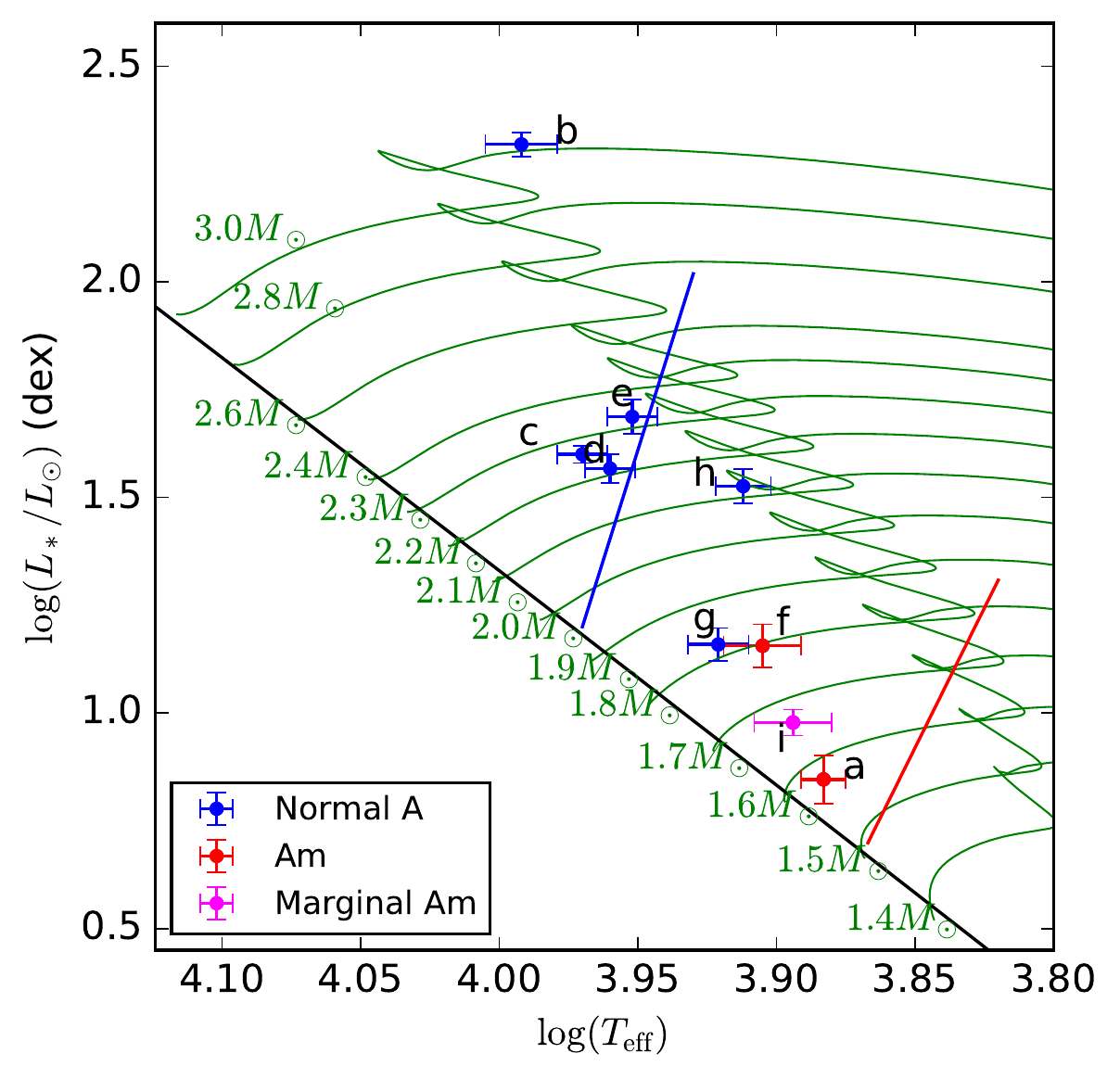}
 \caption{The HR diagram showing evolutionary phase of target stars. The blue, red, and magenta dots are normal A, Am, and marginal Am stars, respectively. The black solid line represents the zero-age main sequence (ZAMS) while the green solid lines are evolutionary tracks for different masses. The blue and red solid lines indicate the blue and red edges of the observational $\delta$\,Scuti instability strip reported by \citet{10.1093/mnras/stz590}.}
 \label{fig:hr}
\end{figure}

Fig.\,\ref{fig:hr} is the HR diagram showing the location of the normal A (blue) and Am (red) stars. With exception of KIC\,4567097, which could be on the red giant phase, all our targets are located in the MS. This was also observed for KIC\,4567097 from MK spectral classification in Sect.\,\ref{class}, where the luminosity class is III (giant), and from the spectral type obtained from SED fitting as discussed in Sect.\,\ref{SEDs}. KIC\,9272082 could be be close to the end of its MS lifetime. All the Am stars lie within the observational $\delta$\,Scuti instability strip determined by \citet{10.1093/mnras/stz590}. With exception of KIC\,7667560 and KIC\,9272082, the normal A-type stars lie outside this instability strip.

\begin{table*}
\centering
\caption{The different columns show: tag (for identification in the HR diagram in Fig.\,\ref{fig:hr}), identification (KIC) number, absolute magnitude ($M_{\rm v}$), bolometric correction (BC), luminosity (\luminosity) and stellar radius ($R$) as estimated from standard relations.}
\label{table:lum}
\begin{tabular}{clcccc}
\hline
\hline
\noalign{\smallskip}
Tag&KIC&$M_{\rm v}$& BC &$ \log (L_\star/L_\odot)$&$R$\\ 
\noalign{\smallskip}
&No.&(mag)&(mag) &(dex)&($R_\odot$)\\
\noalign{\smallskip}
\hline
\noalign{\smallskip}
  a&3459226 & 2.616\,$\pm$\,0.145 & 0.0304\,$\pm$\,0.0038 & 0.846\,$\pm$\,0.056&1.50\,$\pm$\,0.15\\
  b&4567097 & -0.820\,$\pm$\,0.132 & -0.2149\,$\pm$\,0.0610 & 2.319\,$\pm$\,0.028&4.98\,$\pm$\,0.46\\
  c&4818496 & 0.935\,$\pm$\,0.092 & -0.1735\,$\pm$\,0.0410 & 1.600\,$\pm$\,0.020&2.28\,$\pm$\,0.15\\
  d&5524045 & 0.918\,$\pm$\,0.116 & -0.0743\,$\pm$\,0.0303 & 1.567\,$\pm$\,0.034&2.46\,$\pm$\,0.20\\
  e&5650229 & 0.596\,$\pm$\,0.128 & -0.0512\,$\pm$\,0.0276 & 1.687\,$\pm$\,0.040&2.92\,$\pm$\,0.26\\
  f&6266219 & 1.848\,$\pm$\,0.122 & 0.0239\,$\pm$\,0.0024 & 1.156\,$\pm$\,0.050&1.98\,$\pm$\,0.24\\
  g&7667560 & 1.850\,$\pm$\,0.108 & 0.0140\,$\pm$\,0.0130 & 1.159\,$\pm$\,0.038&1.83\,$\pm$\,0.17\\
  h&9272082 & 0.924\,$\pm$\,0.105 & 0.0225\,$\pm$\,0.0047 & 1.526\,$\pm$\,0.040&2.93\,$\pm$\,0.27\\
  i&9349245 & 2.290\,$\pm$\,0.083 & 0.0279\,$\pm$\,0.0084 & 0.978\,$\pm$\,0.030&1.71\,$\pm$\,0.17\\
\hline\end{tabular}
\end{table*}

\section{Conclusion}
\label{concl}

In this paper we present a spectroscopic analysis of a sample of nine objects which are classified in literature as ``hump and spike'' stars. The sample comprises of both normal A and Am stars. The analysis is based on high-resolution spectra obtained with HERMES mounted at the Cassegrain focus of the 1.2-m Mercator telescope located at La Palma, Spain. For each star, we determined the spectral type and obtained fundamental parameters such as the effective temperature \teff, surface gravity \logg, metallicity [M/H], microturbulent velocity $\xi$, projected rotational velocity \vsini\ and radial velocity \vrad. We also performed a detailed individual chemical abundance analysis for each target, obtained their luminosity \luminosity\ and derived an accurate position in the HR diagram.

All the stars are either fast  (KIC\,4818496, KIC\,5524045, KIC\,5650229, and KIC\,7667560) or moderate (KIC\,3459226, KIC\,4567097, KIC\,6266219, KIC\,9272082, and KIC\,9349245)  rotators leading to severe blending of their spectral lines. To overcome this problem, we applied the spectral synthesis method in our analysis using different codes. Of all the stars in our sample, only two stars have high-resolution spectroscopic analysis reported in the literature, i.e. KIC\,5524045 \citep{2015MNRAS.450.2764N} and KIC\,9349245 \citep{catanzaro2015}. For the remaining 7 targets, it was the first time that a detailed spectroscopic analysis was done.

We classified KIC\,4567097 as a giant while the rest are dwarfs. So far, KIC\,6266219 and KIC\,9272082 have been considered as chemically normal and Am, respectively \citep[e.g,][]{2020MNRAS.492.3143T}. Based on our spectral classification analysis and the resulting chemical abundance pattern, we reclassify KIC\,6266219 as an Am star (kA3hA7mF1) while KIC\,9272082 as chemically normal. Therefore, we conclude that our sample contains two Am stars (KIC\,3459226 and KIC\,6266219) and one marginal Am star (KIC\,9349245), and six targets that are considered as non-Am stars (KIC\,4567097, KIC\,4818496, KIC\,5524045, KIC\,5650229, KIC\,7667560, and KIC\,9272082). With exception of the moderate rotators KIC\,4567097 and KIC\,9272082, the remaining 4 non-Am stars are fast rotators (\vsini\,$>120$\,\kms), a property that facilitates mixing against atomic diffusion while minimising the chemical peculiarity. In addition, Fig.\,\ref{vrot_fig}
nicely shows that the Am and marginal Am stars are the slow rotators in the sample, which supports the idea that slow rotation facilitates the creation of chemical peculiarities.  

The availability of HERMES and other spectrographs capable of high-resolution spectroscopy will allow us, in future, to continue the study of the ``hump and spike'' stars. Particularly, we will segregate Am from non-Am stars in the ``hump and spike'' star sample. In the near future, modelling of these stars will shed light on the interaction of chemical peculiarity, rotation, and Rossby waves.

\section*{Acknowledgments}

The Swedish International Development Co-operation Agency (SIDA) through the International Science Program (ISP) of Uppsala University financed this study. The part of work presented here is supported by the Belgo-Indian Network for Astronomy \& Astrophysics (BINA), approved by the International Division, Department of Science and Technology (DST, Govt. of India; DST/INT/Belg/P-02) and the Belgian Federal Science Policy Office (BELSPO, Govt. of Belgium; BL/33/IN12). Based on observations made with the Mercator Telescope, operated on the island of La Palma by the Flemish Community, at the Spanish Observatorio del Roque de los Muchachos of the Instituto de Astrof\'isica de Canarias. This work makes use of \textsc{vosa}, developed under the Spanish Virtual Observatory project supported by the Spanish MINECO through grant AyA2017-84089. VOSA has been partially updated by using funding from the European Union's Horizon 2020 Research and Innovation Programme, under Grant Agreement 776403 (EXOPLANETS-A). We thank Dr. Yves Fr\'{e}mat for kindly providing the \textsc{girfit} code. We acknowledge Professor Vadim Tsymbal for helpful comments and discussions. We acknowledge the anonymous reviewer for his/her careful reading of our manuscript and  many insightful comments and suggestions which improved this paper.


\section*{Data availability}
The data underlying this article are available in the Figshare repository at https://dx.doi.org/10.6084/m9.figshare.13604348.v2.

%
%
%
 \bibliographystyle{mnras}
 \bibliography{ref} 

\begin{thebibliography}{}
\makeatletter
\relax
\def\mn@urlcharsother{\let\do\@makeother \do\$\do\&\do\#\do\^\do\_\do\%\do\~}
\def\mn@doi{\begingroup\mn@urlcharsother \@ifnextchar [ {\mn@doi@}
  {\mn@doi@[]}}
\def\mn@doi@[#1]#2{\def\@tempa{#1}\ifx\@tempa\@empty \href
  {http://dx.doi.org/#2} {doi:#2}\else \href {http://dx.doi.org/#2} {#1}\fi
  \endgroup}
\def\mn@eprint#1#2{\mn@eprint@#1:#2::\@nil}
\def\mn@eprint@arXiv#1{\href {http://arxiv.org/abs/#1} {{\tt arXiv:#1}}}
\def\mn@eprint@dblp#1{\href {http://dblp.uni-trier.de/rec/bibtex/#1.xml}
  {dblp:#1}}
\def\mn@eprint@#1:#2:#3:#4\@nil{\def\@tempa {#1}\def\@tempb {#2}\def\@tempc
  {#3}\ifx \@tempc \@empty \let \@tempc \@tempb \let \@tempb \@tempa \fi \ifx
  \@tempb \@empty \def\@tempb {arXiv}\fi \@ifundefined
  {mn@eprint@\@tempb}{\@tempb:\@tempc}{\expandafter \expandafter \csname
  mn@eprint@\@tempb\endcsname \expandafter{\@tempc}}}

\bibitem[\protect\citeauthoryear{Abt}{Abt}{2009}]{Abt2009}
Abt H.~A.,  2009, \aj, 138, 28

\bibitem[\protect\citeauthoryear{{Alecian}}{{Alecian}}{1996}]{1996A&A...310..872A}
{Alecian} G.,  1996, \aap, \href
  {https://ui.adsabs.harvard.edu/abs/1996A&A...310..872A} {310, 872}

\bibitem[\protect\citeauthoryear{{Andrae} et~al.,}{{Andrae}
  et~al.}{2018}]{2018A&A...616A...8A}
{Andrae} R.,  et~al., 2018, \mn@doi [\aap] {10.1051/0004-6361/201732516}, \href
  {http://adsabs.harvard.edu/abs/2018A%26A...616A...8A} {616, A8}

\bibitem[\protect\citeauthoryear{{Ashoka} et~al.,}{{Ashoka}
  et~al.}{2000}]{2000BASI...28..251A}
{Ashoka} B.~N.,  et~al., 2000, Bulletin of the Astronomical Society of India,
  \href {https://ui.adsabs.harvard.edu/abs/2000BASI...28..251A} {28, 251}

\bibitem[\protect\citeauthoryear{{Asplund}, {Grevesse}, {Sauval}  \&
  {Scott}}{{Asplund} et~al.}{2009}]{2009ARA&A..47..481A}
{Asplund} M.,  {Grevesse} N.,  {Sauval} A.~J.,   {Scott} P.,  2009, \mn@doi
  [\araa] {10.1146/annurev.astro.46.060407.145222}, \href
  {https://ui.adsabs.harvard.edu/abs/2009ARA%26A..47..481A} {47, 481}

\bibitem[\protect\citeauthoryear{{Auri{\`e}re} et~al.,}{{Auri{\`e}re}
  et~al.}{2007}]{2007A&A...475.1053A}
{Auri{\`e}re} M.,  et~al., 2007, \mn@doi [\aap] {10.1051/0004-6361:20078189},
  \href {http://adsabs.harvard.edu/abs/2007A%26A...475.1053A} {475, 1053}

\bibitem[\protect\citeauthoryear{{Baglin}, {Michel}, {Auvergne}  \& {COROT
  Team}}{{Baglin} et~al.}{2006}]{2006ESASP.624E..34B}
{Baglin} A.,  {Michel} E.,  {Auvergne} M.,   {COROT Team} 2006, in Proceedings
  of SOHO 18/GONG 2006/HELAS I, Beyond the spherical Sun. p.~34

\bibitem[\protect\citeauthoryear{{Balona}}{{Balona}}{2013}]{balona13}
{Balona} L.~A.,  2013, \mn@doi [MNRAS] {10.1093/mnras/stt322}, \href
  {http://adsabs.harvard.edu/abs/2013MNRAS.431.2240B} {431, 2240}

\bibitem[\protect\citeauthoryear{{Balona}, {Catanzaro}, {Abedigamba}, {Ripepi}
  \& {Smalley}}{{Balona} et~al.}{2015}]{balona15}
{Balona} L.~A.,  {Catanzaro} G.,  {Abedigamba} O.~P.,  {Ripepi} V.,   {Smalley}
  B.,  2015, \mn@doi [MNRAS] {10.1093/mnras/stv076}, \href
  {http://adsabs.harvard.edu/abs/2015MNRAS.448.1378B} {448, 1378}

\bibitem[\protect\citeauthoryear{{Bayo}, {Rodrigo}, {Barrado Y Navascu{\'e}s},
  {Solano}, {Guti{\'e}rrez}, {Morales-Calder{\'o}n}  \& {Allard}}{{Bayo}
  et~al.}{2008}]{2008A&A...492..277B}
{Bayo} A.,  {Rodrigo} C.,  {Barrado Y Navascu{\'e}s} D.,  {Solano} E.,
  {Guti{\'e}rrez} R.,  {Morales-Calder{\'o}n} M.,   {Allard} F.,  2008, \mn@doi
  [\aap] {10.1051/0004-6361:200810395}, \href
  {http://adsabs.harvard.edu/abs/2008A%26A...492..277B} {492, 277}

\bibitem[\protect\citeauthoryear{{Bertaud}}{{Bertaud}}{1960}]{1960JO.....43..129B}
{Bertaud} C.,  1960, Journal des Observateurs, \href
  {https://ui.adsabs.harvard.edu/abs/1960JO.....43..129B} {43, 129}

\bibitem[\protect\citeauthoryear{{Bianchi} \& {GALEX Team}}{{Bianchi} \& {GALEX
  Team}}{2000}]{2000MmSAI..71.1123B}
{Bianchi} L.,  {GALEX Team} 2000, \memsai, \href
  {http://adsabs.harvard.edu/abs/2000MmSAI..71.1123B} {71, 1123}

\bibitem[\protect\citeauthoryear{{Blanco-Cuaresma}}{{Blanco-Cuaresma}}{2019}]{2019MNRAS.486.2075B}
{Blanco-Cuaresma} S.,  2019, \mn@doi [\mnras] {10.1093/mnras/stz549}, \href
  {https://ui.adsabs.harvard.edu/abs/2019MNRAS.486.2075B} {486, 2075}

\bibitem[\protect\citeauthoryear{{Blanco-Cuaresma}, {Soubiran}, {Heiter}  \&
  {Jofr{\'e}}}{{Blanco-Cuaresma} et~al.}{2014}]{2014A&A...569A.111B}
{Blanco-Cuaresma} S.,  {Soubiran} C.,  {Heiter} U.,   {Jofr{\'e}} P.,  2014,
  \mn@doi [\aap] {10.1051/0004-6361/201423945}, \href
  {https://ui.adsabs.harvard.edu/abs/2014A&A...569A.111B} {569, A111}

\bibitem[\protect\citeauthoryear{{Bohlender}, {Dworetsky}  \&
  {Jomaron}}{{Bohlender} et~al.}{1998}]{1998ApJ...504..533B}
{Bohlender} D.~A.,  {Dworetsky} M.~M.,   {Jomaron} C.~M.,  1998, \mn@doi [\apj]
  {10.1086/306084}, \href
  {https://ui.adsabs.harvard.edu/abs/1998ApJ...504..533B} {504, 533}

\bibitem[\protect\citeauthoryear{{Boltzmann}}{{Boltzmann}}{1884}]{1884AnP...258..291B}
{Boltzmann} L.,  1884, \mn@doi [Annalen der Physik] {10.1002/andp.18842580616},
  \href {https://ui.adsabs.harvard.edu/abs/1884AnP...258..291B} {258, 291}

\bibitem[\protect\citeauthoryear{{Borucki} et~al.,}{{Borucki}
  et~al.}{2010}]{borucki10}
{Borucki} W.~J.,  et~al., 2010, \mn@doi [Science] {10.1126/science.1185402},
  \href {http://adsabs.harvard.edu/abs/2010Sci...327..977B} {327, 977}

\bibitem[\protect\citeauthoryear{{Braithwaite} \& {Spruit}}{{Braithwaite} \&
  {Spruit}}{2004}]{2004Natur.431..819B}
{Braithwaite} J.,  {Spruit} H.~C.,  2004, \mn@doi [\nat] {10.1038/nature02934},
  \href {https://ui.adsabs.harvard.edu/abs/2004Natur.431..819B} {431, 819}

\bibitem[\protect\citeauthoryear{{Brown}, {Latham}, {Everett}  \&
  {Esquerdo}}{{Brown} et~al.}{2011}]{Brown}
{Brown} T.~M.,  {Latham} D.~W.,  {Everett} M.~E.,   {Esquerdo} G.~A.,  2011,
  \mn@doi [AJ] {10.1088/0004-6256/142/4/112}, \href
  {http://adsabs.harvard.edu/abs/2011AJ....142..112B} {142, 112}

\bibitem[\protect\citeauthoryear{{Browning}, {Brun}  \& {Toomre}}{{Browning}
  et~al.}{2004}]{2004ApJ...601..512B}
{Browning} M.~K.,  {Brun} A.~S.,   {Toomre} J.,  2004, \mn@doi [\apj]
  {10.1086/380198}, \href
  {https://ui.adsabs.harvard.edu/abs/2004ApJ...601..512B} {601, 512}

\bibitem[\protect\citeauthoryear{{Cannon} \& {Pickering}}{{Cannon} \&
  {Pickering}}{1993}]{1993yCat.3135....0C}
{Cannon} A.~J.,  {Pickering} E.~C.,  1993, VizieR Online Data Catalog, \href
  {https://ui.adsabs.harvard.edu/abs/1993yCat.3135....0C} {p. III/135A}

\bibitem[\protect\citeauthoryear{{Castelli} \& {Kurucz}}{{Castelli} \&
  {Kurucz}}{2003}]{2003IAUS..210P.A20C}
{Castelli} F.,  {Kurucz} R.~L.,  2003, in {Piskunov} N.,  {Weiss} W.~W.,
  {Gray} D.~F.,  eds,  IAU Symposium Vol. 210, Modelling of Stellar
  Atmospheres. p.~A20

\bibitem[\protect\citeauthoryear{{Castelli}, {Gratton}  \& {Kurucz}}{{Castelli}
  et~al.}{1997}]{1997A&A...318..841C}
{Castelli} F.,  {Gratton} R.~G.,   {Kurucz} R.~L.,  1997, \aap, \href
  {http://adsabs.harvard.edu/abs/1997A%26A...318..841C} {318, 841}

\bibitem[\protect\citeauthoryear{{Catanzaro} et~al.,}{{Catanzaro}
  et~al.}{2015}]{catanzaro2015}
{Catanzaro} G.,  et~al., 2015, \mn@doi [\mnras] {10.1093/mnras/stv952}, \href
  {http://adsabs.harvard.edu/abs/2015MNRAS.451..184C} {451, 184}

\bibitem[\protect\citeauthoryear{{Chambers} et~al.,}{{Chambers}
  et~al.}{2016}]{2016arXiv161205560C}
{Chambers} K.~C.,  et~al., 2016, preprint, \href
  {http://adsabs.harvard.edu/abs/2016arXiv161205560C} {} (\mn@eprint {arXiv}
  {1612.05560})

\bibitem[\protect\citeauthoryear{{Claverie}, {Isaak}, {McLeod}, {van der Raay}
  \& {Roca Cortes}}{{Claverie} et~al.}{1981}]{1981Natur.293..443C}
{Claverie} A.,  {Isaak} G.~R.,  {McLeod} C.~P.,  {van der Raay} H.~B.,   {Roca
  Cortes} T.,  1981, \mn@doi [\nat] {10.1038/293443a0}, \href
  {https://ui.adsabs.harvard.edu/abs/1981Natur.293..443C} {293, 443}

\bibitem[\protect\citeauthoryear{{Conti}}{{Conti}}{1970}]{1970PASP...82..781C}
{Conti} P.~S.,  1970, \mn@doi [PASP] {10.1086/128965}, \href
  {https://ui.adsabs.harvard.edu/abs/1970PASP...82..781C} {82, 781}

\bibitem[\protect\citeauthoryear{{Costa}, {Girardi}, {Bressan}, {Marigo},
  {Rodrigues}, {Chen}, {Lanza}  \& {Goudfrooij}}{{Costa}
  et~al.}{2019}]{2019MNRAS.485.4641C}
{Costa} G.,  {Girardi} L.,  {Bressan} A.,  {Marigo} P.,  {Rodrigues} T.~S.,
  {Chen} Y.,  {Lanza} A.,   {Goudfrooij} P.,  2019, \mn@doi [\mnras]
  {10.1093/mnras/stz728}, \href
  {https://ui.adsabs.harvard.edu/abs/2019MNRAS.485.4641C} {485, 4641}

\bibitem[\protect\citeauthoryear{{Cui} et~al.,}{{Cui}
  et~al.}{2012}]{2012RAA....12.1197C}
{Cui} X.-Q.,  et~al., 2012, \mn@doi [Research in Astronomy and Astrophysics]
  {10.1088/1674-4527/12/9/003}, \href
  {https://ui.adsabs.harvard.edu/abs/2012RAA....12.1197C} {12, 1197}

\bibitem[\protect\citeauthoryear{{Cutri} et~al.,}{{Cutri}
  et~al.}{2003}]{2003yCat.2246....0C}
{Cutri} R.~M.,  et~al., 2003, VizieR Online Data Catalog, \href
  {https://ui.adsabs.harvard.edu/abs/2003yCat.2246....0C} {p. II/246}

\bibitem[\protect\citeauthoryear{{Dziembowski}, {Krolikowska}  \&
  {Kosovichev}}{{Dziembowski} et~al.}{1988}]{1988AcA....38...61D}
{Dziembowski} W.,  {Krolikowska} M.,   {Kosovichev} A.,  1988, \actaa, \href
  {https://ui.adsabs.harvard.edu/abs/1988AcA....38...61D} {38, 61}

\bibitem[\protect\citeauthoryear{Eisner et~al.,}{Eisner
  et~al.}{2020}]{10.1093/mnras/staa138}
Eisner N.~L.,  et~al., 2020, \mn@doi [\mnras] {10.1093/mnras/staa138}, 494, 750

\bibitem[\protect\citeauthoryear{{Floquet}}{{Floquet}}{1970}]{1970A&AS....1....1F}
{Floquet} M.,  1970, \aaps, \href
  {https://ui.adsabs.harvard.edu/abs/1970A&AS....1....1F} {1, 1}

\bibitem[\protect\citeauthoryear{{Flower}}{{Flower}}{1996}]{1996ApJ...469..355F}
{Flower} P.~J.,  1996, \mn@doi [ApJ] {10.1086/177785}, \href
  {http://adsabs.harvard.edu/abs/1996ApJ...469..355F} {469, 355}

\bibitem[\protect\citeauthoryear{{Fossati}, {Bagnulo}, {Landstreet}, {Wade},
  {Kochukhov}, {Monier}, {Weiss}  \& {Gebran}}{{Fossati}
  et~al.}{2008}]{2008A&A...483..891F}
{Fossati} L.,  {Bagnulo} S.,  {Landstreet} J.,  {Wade} G.,  {Kochukhov} O.,
  {Monier} R.,  {Weiss} W.,   {Gebran} M.,  2008, \mn@doi [\aap]
  {10.1051/0004-6361:200809467}, \href
  {https://ui.adsabs.harvard.edu/abs/2008A&A...483..891F} {483, 891}

\bibitem[\protect\citeauthoryear{{Frasca} et~al.,}{{Frasca}
  et~al.}{2016}]{2016A&A...594A..39F}
{Frasca} A.,  et~al., 2016, \mn@doi [\aap] {10.1051/0004-6361/201628337}, \href
  {https://ui.adsabs.harvard.edu/abs/2016A&A...594A..39F} {594, A39}

\bibitem[\protect\citeauthoryear{{Fr{\'e}mat}, {Neiner}, {Hubert}, {Floquet},
  {Zorec}, {Janot-Pacheco}  \& {Renan de Medeiros}}{{Fr{\'e}mat}
  et~al.}{2006}]{2006A&A...451.1053F}
{Fr{\'e}mat} Y.,  {Neiner} C.,  {Hubert} A.~M.,  {Floquet} M.,  {Zorec} J.,
  {Janot-Pacheco} E.,   {Renan de Medeiros} J.,  2006, \mn@doi [\aap]
  {10.1051/0004-6361:20053305}, \href
  {https://ui.adsabs.harvard.edu/abs/2006A&A...451.1053F} {451, 1053}

\bibitem[\protect\citeauthoryear{{Gaia Collaboration}}{{Gaia
  Collaboration}}{2018}]{2018yCat.1345....0G}
{Gaia Collaboration} 2018, VizieR Online Data Catalog, \href
  {https://ui.adsabs.harvard.edu/\#abs/2018yCat.1345....0G} {p. I/345}

\bibitem[\protect\citeauthoryear{{Gandolfi} et~al.,}{{Gandolfi}
  et~al.}{2018}]{2018A&A...619L..10G}
{Gandolfi} D.,  et~al., 2018, \mn@doi [\aap] {10.1051/0004-6361/201834289},
  \href {https://ui.adsabs.harvard.edu/abs/2018A&A...619L..10G} {619, L10}

\bibitem[\protect\citeauthoryear{{Gebran} \& {Monier}}{{Gebran} \&
  {Monier}}{2008}]{2008A&A...483..567G}
{Gebran} M.,  {Monier} R.,  2008, \mn@doi [\aap] {10.1051/0004-6361:20079271},
  \href {https://ui.adsabs.harvard.edu/abs/2008A&A...483..567G} {483, 567}

\bibitem[\protect\citeauthoryear{{Gebran}, {Monier}  \& {Richard}}{{Gebran}
  et~al.}{2008}]{2008A&A...479..189G}
{Gebran} M.,  {Monier} R.,   {Richard} O.,  2008, \mn@doi [\aap]
  {10.1051/0004-6361:20078807}, \href
  {https://ui.adsabs.harvard.edu/abs/2008A&A...479..189G} {479, 189}

\bibitem[\protect\citeauthoryear{{Gebran}, {Vick}, {Monier}  \&
  {Fossati}}{{Gebran} et~al.}{2010}]{2010A&A...523A..71G}
{Gebran} M.,  {Vick} M.,  {Monier} R.,   {Fossati} L.,  2010, \mn@doi [\aap]
  {10.1051/0004-6361/200913273}, \href
  {https://ui.adsabs.harvard.edu/abs/2010A&A...523A..71G} {523, A71}

\bibitem[\protect\citeauthoryear{{Gebran}, {Monier}, {Royer}, {Lobel}  \&
  {Blomme}}{{Gebran} et~al.}{2014}]{gebran14}
{Gebran} M.,  {Monier} R.,  {Royer} F.,  {Lobel} A.,   {Blomme} R.,  2014, in
  {Mathys} G.,  {Griffin} E.~R.,  {Kochukhov} O.,  {Monier} R.,   {Wahlgren}
  G.~M.,  eds, Putting A Stars into Context: Evolution, Environment, and
  Related Stars. pp 193--198 (\mn@eprint {arXiv} {1312.0442})

\bibitem[\protect\citeauthoryear{{Gehrels}, {Spergel}  \& {WFIRST SDT
  Project}}{{Gehrels} et~al.}{2015}]{Gehrels_2015}
{Gehrels} N.,  {Spergel} D.,   {WFIRST SDT Project} 2015, \mn@doi [J. Phys.:
  Conf. Ser.] {10.1088/1742-6596/610/1/012007}, 610, 012007

\bibitem[\protect\citeauthoryear{{Gontcharov}}{{Gontcharov}}{2006}]{2006AstL...32..759G}
{Gontcharov} G.~A.,  2006, \mn@doi [Astron. Lett.] {10.1134/S1063773706110065},
  \href {https://ui.adsabs.harvard.edu/abs/2006AstL...32..759G} {32, 759}

\bibitem[\protect\citeauthoryear{{Gray} \& {Corbally}}{{Gray} \&
  {Corbally}}{2009}]{2009ssc..book.....G}
{Gray} R.~O.,  {Corbally} Christopher J.,  2009, {Stellar Spectral
  Classification}.
Princeton University Press

\bibitem[\protect\citeauthoryear{{Gray} \& {Corbally}}{{Gray} \&
  {Corbally}}{2014}]{2014AJ....147...80G}
{Gray} R.~O.,  {Corbally} C.~J.,  2014, \mn@doi [\aj]
  {10.1088/0004-6256/147/4/80}, \href
  {https://ui.adsabs.harvard.edu/abs/2014AJ....147...80G} {147, 80}

\bibitem[\protect\citeauthoryear{{Gray}, {Corbally}, {Garrison}, {McFadden}  \&
  {Robinson}}{{Gray} et~al.}{2003}]{2003AJ....126.2048G}
{Gray} R.~O.,  {Corbally} C.~J.,  {Garrison} R.~F.,  {McFadden} M.~T.,
  {Robinson} P.~E.,  2003, \mn@doi [\aj] {10.1086/378365}, \href
  {https://ui.adsabs.harvard.edu/abs/2003AJ....126.2048G} {126, 2048}

\bibitem[\protect\citeauthoryear{{Gray} et~al.,}{{Gray}
  et~al.}{2016}]{2016AJ....151...13G}
{Gray} R.~O.,  et~al., 2016, \mn@doi [\aj] {10.3847/0004-6256/151/1/13}, \href
  {http://adsabs.harvard.edu/abs/2016AJ....151...13G} {151, 13}

\bibitem[\protect\citeauthoryear{{Green}}{{Green}}{2018}]{bayestar}
{Green} G.~M.,  2018, \mn@doi [JOSS] {https://doi.org/10.21105/joss.00695}, 3,
  695

\bibitem[\protect\citeauthoryear{{Green}, {Schlafly}, {Zucker}, {Speagle}  \&
  {Finkbeiner}}{{Green} et~al.}{2019}]{2019arXiv190502734G}
{Green} G.~M.,  {Schlafly} E.,  {Zucker} C.,  {Speagle} J.~S.,   {Finkbeiner}
  D.,  2019, \mn@doi [\apj] {10.3847/1538-4357/ab5362}, \href
  {https://ui.adsabs.harvard.edu/abs/2019ApJ...887...93G} {887, 93}

\bibitem[\protect\citeauthoryear{{Hauck} \& {Mermilliod}}{{Hauck} \&
  {Mermilliod}}{1998}]{1998A&AS..129..431H}
{Hauck} B.,  {Mermilliod} M.,  1998, \mn@doi [\aaps] {10.1051/aas:1998195},
  \href {http://adsabs.harvard.edu/abs/1998A%26AS..129..431H} {129, 431}

\bibitem[\protect\citeauthoryear{{Herdin}, {Paunzen}  \& {Netopil}}{{Herdin}
  et~al.}{2016}]{2016A&A...585A..67H}
{Herdin} A.,  {Paunzen} E.,   {Netopil} M.,  2016, \mn@doi [\aap]
  {10.1051/0004-6361/201527390}, \href
  {https://ui.adsabs.harvard.edu/abs/2016A&A...585A..67H} {585, A67}

\bibitem[\protect\citeauthoryear{{H{\o}g} et~al.,}{{H{\o}g}
  et~al.}{2000a}]{2000A&A...355L..27H}
{H{\o}g} E.,  et~al., 2000a, \aap, \href
  {https://ui.adsabs.harvard.edu/abs/2000A&A...355L..27H} {355, L27}

\bibitem[\protect\citeauthoryear{{H{\o}g} et~al.,}{{H{\o}g}
  et~al.}{2000b}]{2000A&A...357..367H}
{H{\o}g} E.,  et~al., 2000b, \aap, \href
  {http://adsabs.harvard.edu/abs/2000A%26A...357..367H} {357, 367}

\bibitem[\protect\citeauthoryear{{Howard} et~al.,}{{Howard}
  et~al.}{2013}]{2013Natur.503..381H}
{Howard} A.~W.,  et~al., 2013, \mn@doi [\nat] {10.1038/nature12767}, \href
  {https://ui.adsabs.harvard.edu/abs/2013Natur.503..381H} {503, 381}

\bibitem[\protect\citeauthoryear{{Howell} et~al.,}{{Howell}
  et~al.}{2014}]{2014PASP..126..398H}
{Howell} S.~B.,  et~al., 2014, \mn@doi [\pasp] {10.1086/676406}, \href
  {https://ui.adsabs.harvard.edu/abs/2014PASP..126..398H} {126, 398}

\bibitem[\protect\citeauthoryear{{Hui-Bon-Hoa}}{{Hui-Bon-Hoa}}{2000}]{hui}
{Hui-Bon-Hoa} A.,  2000, \mn@doi [A\&AS] {10.1051/aas:2000207}, \href
  {http://adsabs.harvard.edu/abs/2000A%26AS..144..203H} {144, 203}

\bibitem[\protect\citeauthoryear{{Johnson} \& {Morgan}}{{Johnson} \&
  {Morgan}}{1953}]{1953ApJ...117..313J}
{Johnson} H.~L.,  {Morgan} W.~W.,  1953, \mn@doi [\apj] {10.1086/145697}, \href
  {https://ui.adsabs.harvard.edu/abs/1953ApJ...117..313J} {117, 313}

\bibitem[\protect\citeauthoryear{{Jones} \& {Wolff}}{{Jones} \&
  {Wolff}}{1974}]{1974PASP...86...67J}
{Jones} T.~J.,  {Wolff} S.~C.,  1974, \mn@doi [\pasp] {10.1086/129561}, \href
  {https://ui.adsabs.harvard.edu/abs/1974PASP...86...67J} {86, 67}

\bibitem[\protect\citeauthoryear{{Joshi} et~al.,}{{Joshi}
  et~al.}{2003}]{2003MNRAS.344..431J}
{Joshi} S.,  et~al., 2003, \mn@doi [\mnras] {10.1046/j.1365-8711.2003.06823.x},
  \href {http://cdsads.u-strasbg.fr/abs/2003MNRAS.344..431J} {344, 431}

\bibitem[\protect\citeauthoryear{{Joshi}, {Mary}, {Martinez}, {Kurtz},
  {Girish}, {Seetha}, {Sagar}  \& {Ashoka}}{{Joshi}
  et~al.}{2006}]{2006A&A...455..303J}
{Joshi} S.,  {Mary} D.~L.,  {Martinez} P.,  {Kurtz} D.~W.,  {Girish} V.,
  {Seetha} S.,  {Sagar} R.,   {Ashoka} B.~N.,  2006, \mn@doi [\aap]
  {10.1051/0004-6361:20064970}, \href
  {https://ui.adsabs.harvard.edu/abs/2006A%26A...455..303J} {455, 303}

\bibitem[\protect\citeauthoryear{{Joshi}, {Mary}, {Chakradhari}, {Tiwari}  \&
  {Billaud}}{{Joshi} et~al.}{2009}]{2009A&A...507.1763J}
{Joshi} S.,  {Mary} D.~L.,  {Chakradhari} N.~K.,  {Tiwari} S.~K.,   {Billaud}
  C.,  2009, \mn@doi [\aap] {10.1051/0004-6361/200912382}, \href
  {https://ui.adsabs.harvard.edu/abs/2009A%26A...507.1763J} {507, 1763}

\bibitem[\protect\citeauthoryear{{Joshi}, {Ryabchikova}, {Kochukhov},
  {Sachkov}, {Tiwari}, {Chakradhari}  \& {Piskunov}}{{Joshi}
  et~al.}{2010}]{2010MNRAS.401.1299J}
{Joshi} S.,  {Ryabchikova} T.,  {Kochukhov} O.,  {Sachkov} M.,  {Tiwari} S.~K.,
   {Chakradhari} N.~K.,   {Piskunov} N.,  2010, \mn@doi [\mnras]
  {10.1111/j.1365-2966.2009.15725.x}, \href
  {http://cdsads.u-strasbg.fr/abs/2010MNRAS.401.1299J} {401, 1299}

\bibitem[\protect\citeauthoryear{{Joshi} et~al.,}{{Joshi}
  et~al.}{2012}]{2012MNRAS.424.2002J}
{Joshi} S.,  et~al., 2012, \mn@doi [\mnras] {10.1111/j.1365-2966.2012.21340.x},
  \href {http://cdsads.u-strasbg.fr/abs/2012MNRAS.424.2002J} {424, 2002}

\bibitem[\protect\citeauthoryear{{Joshi} et~al.,}{{Joshi}
  et~al.}{2016}]{2016A&A...590A.116J}
{Joshi} S.,  et~al., 2016, \mn@doi [\aap] {10.1051/0004-6361/201527242}, \href
  {https://ui.adsabs.harvard.edu/abs/2016A%26A...590A.116J} {590, A116}

\bibitem[\protect\citeauthoryear{{Joshi}, {Semenko}, {Moiseeva}, {Sharma},
  {Joshi}, {Sachkov}, {Singh}  \& {Yerra}}{{Joshi}
  et~al.}{2017}]{2017MNRAS.467..633J}
{Joshi} S.,  {Semenko} E.,  {Moiseeva} A.,  {Sharma} K.,  {Joshi} Y.~C.,
  {Sachkov} M.,  {Singh} H.~P.,   {Yerra} B.~K.,  2017, \mn@doi [\mnras]
  {10.1093/mnras/stx087}, \href
  {http://adsabs.harvard.edu/abs/2017MNRAS.467..633J} {467, 633}

\bibitem[\protect\citeauthoryear{{Kahraman Ali{\c{c}}avu{\textcommabelow s}}
  et~al.,}{{Kahraman Ali{\c{c}}avu{\textcommabelow s}}
  et~al.}{2016}]{2016MNRAS.458.2307K}
{Kahraman Ali{\c{c}}avu{\textcommabelow s}} F.,  et~al., 2016, \mn@doi [\mnras]
  {10.1093/mnras/stw393}, \href
  {https://ui.adsabs.harvard.edu/abs/2016MNRAS.458.2307K} {458, 2307}

\bibitem[\protect\citeauthoryear{{Kaler}}{{Kaler}}{1989}]{1989ssis.book.....K}
{Kaler} J.~B.,  1989, {Stars and their spectra. an introduction to spectral
  sequence}.
Cambridge University Press

\bibitem[\protect\citeauthoryear{{Kassounian}, {Gebran}, {Paletou}  \&
  {Watson}}{{Kassounian} et~al.}{2019}]{2019OAst...28...68K}
{Kassounian} S.,  {Gebran} M.,  {Paletou} F.,   {Watson} V.,  2019, \mn@doi
  [OAst] {10.1515/astro-2019-0006}, \href
  {https://ui.adsabs.harvard.edu/abs/2019OAst...28...68K} {28, 68}

\bibitem[\protect\citeauthoryear{{Kesseli}, {West}, {Veyette}, {Harrison},
  {Feldman}  \& {Bochanski}}{{Kesseli} et~al.}{2017}]{2017ApJS..230...16K}
{Kesseli} A.~Y.,  {West} A.~A.,  {Veyette} M.,  {Harrison} B.,  {Feldman} D.,
  {Bochanski} J.~J.,  2017, \mn@doi [\apjs] {10.3847/1538-4365/aa656d}, \href
  {http://adsabs.harvard.edu/abs/2017ApJS..230...16K} {230, 16}

\bibitem[\protect\citeauthoryear{{Khokhlova}}{{Khokhlova}}{1981}]{khokh}
{Khokhlova} V.~L.,  1981, in Chemically peculiar stars of the upper main
  sequence. Universit\'{e} de Li\`{e}ge, pp 457--463

\bibitem[\protect\citeauthoryear{{Kochukhov}}{{Kochukhov}}{2018}]{2018ascl.soft05015K}
{Kochukhov} O.,  2018, {BinMag: Widget for comparing stellar observed with
  theoretical spectra} (\mn@eprint {ascl} {1805.015})

\bibitem[\protect\citeauthoryear{{Kron}}{{Kron}}{1947}]{1947PASP...59..261K}
{Kron} G.~E.,  1947, \mn@doi [\pasp] {10.1086/125964}, \href
  {https://ui.adsabs.harvard.edu/abs/1947PASP...59..261K} {59, 261}

\bibitem[\protect\citeauthoryear{{Kupka}, {Piskunov}, {Ryabchikova}, {Stempels}
   \& {Weiss}}{{Kupka} et~al.}{1999}]{1999A&AS..138..119K}
{Kupka} F.,  {Piskunov} N.,  {Ryabchikova} T.~A.,  {Stempels} H.~C.,   {Weiss}
  W.~W.,  1999, \mn@doi [\aaps] {10.1051/aas:1999267}, \href
  {https://ui.adsabs.harvard.edu/abs/1999A&AS..138..119K} {138, 119}

\bibitem[\protect\citeauthoryear{{Kurtz} \& {Martinez}}{{Kurtz} \&
  {Martinez}}{2000}]{2000BaltA...9..253K}
{Kurtz} D.~W.,  {Martinez} P.,  2000, \mn@doi [BaltA]
  {10.1515/astro-2000-0209}, \href
  {https://ui.adsabs.harvard.edu/abs/2000BaltA...9..253K} {9, 253}

\bibitem[\protect\citeauthoryear{{LaSala}}{{LaSala}}{1994}]{1994ASPC...60..312L}
{LaSala} J.,  1994, in {Corbally} C.~J.,  {Gray} R.~O.,   {Garrison} R.~F.,
  eds,  Astronomical Society of the Pacific Conference Series Vol. 60, The MK
  Process at 50 Years: A Powerful Tool for Astrophysical Insight. p.~312

\bibitem[\protect\citeauthoryear{{Macrae}}{{Macrae}}{1952}]{1952ApJ...116..592M}
{Macrae} D.~A.,  1952, \mn@doi [\apj] {10.1086/145652}, \href
  {https://ui.adsabs.harvard.edu/abs/1952ApJ...116..592M} {116, 592}

\bibitem[\protect\citeauthoryear{{Ma{\'\i}z Apell{\'a}niz} \&
  {Weiler}}{{Ma{\'\i}z Apell{\'a}niz} \& {Weiler}}{2018}]{2018A&A...619A.180M}
{Ma{\'\i}z Apell{\'a}niz} J.,  {Weiler} M.,  2018, \mn@doi [\aap]
  {10.1051/0004-6361/201834051}, \href
  {https://ui.adsabs.harvard.edu/abs/2018A&A...619A.180M} {619, A180}

\bibitem[\protect\citeauthoryear{{Martinez} et~al.,}{{Martinez}
  et~al.}{2001}]{2001AA...371.1048M}
{Martinez} P.,  et~al., 2001, \mn@doi [\aap] {10.1051/0004-6361:20010432},
  \href {http://cdsads.u-strasbg.fr/abs/2001A%26A...371.1048M} {371, 1048}

\bibitem[\protect\citeauthoryear{{Masana}, {Jordi}  \& {Ribas}}{{Masana}
  et~al.}{2006}]{2006A&A...450..735M}
{Masana} E.,  {Jordi} C.,   {Ribas} I.,  2006, \mn@doi [A\&A]
  {10.1051/0004-6361:20054021}, \href
  {https://ui.adsabs.harvard.edu/\#abs/2006A&A...450..735M} {450, 735}

\bibitem[\protect\citeauthoryear{{Mathur} et~al.,}{{Mathur}
  et~al.}{2017}]{2017ApJS..229...30M}
{Mathur} S.,  et~al., 2017, \mn@doi [ApJS] {10.3847/1538-4365/229/2/30}, \href
  {https://ui.adsabs.harvard.edu/abs/2017ApJS..229...30M} {229, 30}

\bibitem[\protect\citeauthoryear{{McDonald}, {Zijlstra}  \& {Boyer}}{{McDonald}
  et~al.}{2012}]{2012MNRAS.427..343M}
{McDonald} I.,  {Zijlstra} A.~A.,   {Boyer} M.~L.,  2012, \mn@doi [\mnras]
  {10.1111/j.1365-2966.2012.21873.x}, \href
  {https://ui.adsabs.harvard.edu/abs/2012MNRAS.427..343M} {427, 343}

\bibitem[\protect\citeauthoryear{{Mermilliod}, {Mermilliod}  \&
  {Hauck}}{{Mermilliod} et~al.}{1997}]{1997A&AS..124..349M}
{Mermilliod} J.-C.,  {Mermilliod} M.,   {Hauck} B.,  1997, \mn@doi [A\&AS]
  {10.1051/aas:1997197}, \href
  {http://adsabs.harvard.edu/abs/1997A%26AS..124..349M} {124, 349}

\bibitem[\protect\citeauthoryear{{Michaud}}{{Michaud}}{1970}]{1970ApJ...160..641M}
{Michaud} G.,  1970, \mn@doi [ApJ] {10.1086/150459}, \href
  {http://adsabs.harvard.edu/abs/1970ApJ...160..641M} {160, 641}

\bibitem[\protect\citeauthoryear{{Michaud} \& {Charland}}{{Michaud} \&
  {Charland}}{1986}]{1986ApJ...311..326M}
{Michaud} G.,  {Charland} Y.,  1986, \mn@doi [\apj] {10.1086/164774}, \href
  {https://ui.adsabs.harvard.edu/abs/1986ApJ...311..326M} {311, 326}

\bibitem[\protect\citeauthoryear{{Michaud}, {Tarasick}, {Charland}  \&
  {Pelletier}}{{Michaud} et~al.}{1983}]{1983ApJ...269..239M}
{Michaud} G.,  {Tarasick} D.,  {Charland} Y.,   {Pelletier} C.,  1983, \mn@doi
  [\apj] {10.1086/161034}, \href
  {https://ui.adsabs.harvard.edu/abs/1983ApJ...269..239M} {269, 239}

\bibitem[\protect\citeauthoryear{{Michaud}, {Richard}, {Richer}  \&
  {VandenBerg}}{{Michaud} et~al.}{2004}]{2004ApJ...606..452M}
{Michaud} G.,  {Richard} O.,  {Richer} J.,   {VandenBerg} D.~A.,  2004, \mn@doi
  [\apj] {10.1086/383001}, \href
  {https://ui.adsabs.harvard.edu/abs/2004ApJ...606..452M} {606, 452}

\bibitem[\protect\citeauthoryear{{Montambaux}}{{Montambaux}}{2018}]{Montambaux2018}
{Montambaux} G.,  2018, \mn@doi [Foundations of Physics]
  {10.1007/s10701-018-0153-4}, 48, 395–410

\bibitem[\protect\citeauthoryear{{Moon}}{{Moon}}{1985}]{moon1985stellar}
{Moon} T.,  1985, Comm. Univ. London Obs., 78, 1

\bibitem[\protect\citeauthoryear{{Moon} \& {Dworetsky}}{{Moon} \&
  {Dworetsky}}{1985}]{1985MNRAS.217..305M}
{Moon} T.~T.,  {Dworetsky} M.~M.,  1985, \mn@doi [\mnras]
  {10.1093/mnras/217.2.305}, \href
  {http://adsabs.harvard.edu/abs/1985MNRAS.217..305M} {217, 305}

\bibitem[\protect\citeauthoryear{{Morgan}, {Keenan}  \& {Kellman}}{{Morgan}
  et~al.}{1943}]{1943assw.book.....M}
{Morgan} W.~W.,  {Keenan} P.~C.,   {Kellman} E.,  1943, {An atlas of stellar
  spectra, with an outline of spectral classification}.
University of Chicago press

\bibitem[\protect\citeauthoryear{{Mosser}, {Baudin}, {Lanza}, {Hulot},
  {Catala}, {Baglin}  \& {Auvergne}}{{Mosser}
  et~al.}{2009}]{2009A&A...506..245M}
{Mosser} B.,  {Baudin} F.,  {Lanza} A.~F.,  {Hulot} J.~C.,  {Catala} C.,
  {Baglin} A.,   {Auvergne} M.,  2009, \mn@doi [\aap]
  {10.1051/0004-6361/200911942}, \href
  {https://ui.adsabs.harvard.edu/abs/2009A%26A...506..245M} {506, 245}

\bibitem[\protect\citeauthoryear{{Murphy}}{{Murphy}}{2014}]{2014PhDT.......131M}
{Murphy} S.~J.,  2014, PhD thesis, Jeremiah Horrocks Institute, University of
  Central Lancashire, Preston, UK <EMAIL>murphy@physics.usyd.edu.au</EMAIL>

\bibitem[\protect\citeauthoryear{Murphy, Hey, Van Reeth  \& Bedding}{Murphy
  et~al.}{2019}]{10.1093/mnras/stz590}
Murphy S.~J.,  Hey D.,  Van Reeth T.,   Bedding T.~R.,  2019, \mn@doi [\mnras]
  {10.1093/mnras/stz590}, 485, 2380

\bibitem[\protect\citeauthoryear{{Napiwotzki}, {Schoenberner}  \&
  {Wenske}}{{Napiwotzki} et~al.}{1993}]{1993A&A...268..653N}
{Napiwotzki} R.,  {Schoenberner} D.,   {Wenske} V.,  1993, \aap, \href
  {http://adsabs.harvard.edu/abs/1993A%26A...268..653N} {268, 653}

\bibitem[\protect\citeauthoryear{{Niemczura} et~al.,}{{Niemczura}
  et~al.}{2015}]{2015MNRAS.450.2764N}
{Niemczura} E.,  et~al., 2015, \mn@doi [\mnras] {10.1093/mnras/stv528}, \href
  {https://ui.adsabs.harvard.edu/abs/2015MNRAS.450.2764N} {450, 2764}

\bibitem[\protect\citeauthoryear{{Niemczura} et~al.,}{{Niemczura}
  et~al.}{2017}]{2017MNRAS.470.2870N}
{Niemczura} E.,  et~al., 2017, \mn@doi [MNRAS] {10.1093/mnras/stx1256}, \href
  {https://ui.adsabs.harvard.edu/abs/2017MNRAS.470.2870N} {470, 2870}

\bibitem[\protect\citeauthoryear{{Noyes}, {Hartmann}, {Baliunas}, {Duncan}  \&
  {Vaughan}}{{Noyes} et~al.}{1984}]{1984ApJ...279..763N}
{Noyes} R.~W.,  {Hartmann} L.~W.,  {Baliunas} S.~L.,  {Duncan} D.~K.,
  {Vaughan} A.~H.,  1984, \mn@doi [\apj] {10.1086/161945}, \href
  {https://ui.adsabs.harvard.edu/abs/1984ApJ...279..763N} {279, 763}

\bibitem[\protect\citeauthoryear{{Oja}}{{Oja}}{1985}]{1985A&AS...59..461O}
{Oja} T.,  1985, \aaps, \href
  {https://ui.adsabs.harvard.edu/abs/1985A&AS...59..461O} {59, 461}

\bibitem[\protect\citeauthoryear{{Paul}, {Greenberger}, {Stenholm}  \&
  {Schleich}}{{Paul} et~al.}{2015}]{2015PhST..165a4027P}
{Paul} H.,  {Greenberger} D.~M.,  {Stenholm} S.~T.,   {Schleich} W.~P.,  2015,
  \mn@doi [Physica Scripta Volume T] {10.1088/0031-8949/2015/T165/014027},
  \href {https://ui.adsabs.harvard.edu/abs/2015PhST..165a4027P} {165, 014027}

\bibitem[\protect\citeauthoryear{{Preston}}{{Preston}}{1974}]{1974ARA&A..12..257P}
{Preston} G.~W.,  1974, \mn@doi [\araa] {10.1146/annurev.aa.12.090174.001353},
  \href {http://adsabs.harvard.edu/abs/1974ARA%26A..12..257P} {12, 257}

\bibitem[\protect\citeauthoryear{{Przybilla}, {Butler}, {Becker}, {Kudritzki}
  \& {Venn}}{{Przybilla} et~al.}{2000}]{2000A&A...359.1085P}
{Przybilla} N.,  {Butler} K.,  {Becker} S.~R.,  {Kudritzki} R.~P.,   {Venn}
  K.~A.,  2000, \aap, \href
  {https://ui.adsabs.harvard.edu/abs/2000A&A...359.1085P} {359, 1085}

\bibitem[\protect\citeauthoryear{{Ram{\'\i}rez} \&
  {Mel{\'e}ndez}}{{Ram{\'\i}rez} \& {Mel{\'e}ndez}}{2005}]{2005ApJ...626..465R}
{Ram{\'\i}rez} I.,  {Mel{\'e}ndez} J.,  2005, \mn@doi [\apj] {10.1086/430102},
  \href {https://ui.adsabs.harvard.edu/abs/2005ApJ...626..465R} {626, 465}

\bibitem[\protect\citeauthoryear{{Raskin} et~al.,}{{Raskin}
  et~al.}{2011}]{2011A&A...526A..69R}
{Raskin} G.,  et~al., 2011, \mn@doi [\aap] {10.1051/0004-6361/201015435}, \href
  {http://adsabs.harvard.edu/abs/2011A%26A...526A..69R} {526, A69}

\bibitem[\protect\citeauthoryear{{Rauer} et~al.,}{{Rauer}
  et~al.}{2014}]{2014ExA....38..249R}
{Rauer} H.,  et~al., 2014, \mn@doi [Exp. Astron.] {10.1007/s10686-014-9383-4},
  \href {https://ui.adsabs.harvard.edu/abs/2014ExA....38..249R} {38, 249}

\bibitem[\protect\citeauthoryear{{Renson} \& {Manfroid}}{{Renson} \&
  {Manfroid}}{2009}]{2009A&A...498..961R}
{Renson} P.,  {Manfroid} J.,  2009, \mn@doi [\aap]
  {10.1051/0004-6361/200810788}, \href
  {https://ui.adsabs.harvard.edu/abs/2009A&A...498..961R} {498, 961}

\bibitem[\protect\citeauthoryear{{Richard}, {Michaud}  \& {Richer}}{{Richard}
  et~al.}{2001}]{2001ApJ...558..377R}
{Richard} O.,  {Michaud} G.,   {Richer} J.,  2001, \mn@doi [\apj]
  {10.1086/322264}, \href
  {https://ui.adsabs.harvard.edu/abs/2001ApJ...558..377R} {558, 377}

\bibitem[\protect\citeauthoryear{{Richer}, {Michaud}  \& {Turcotte}}{{Richer}
  et~al.}{2000}]{2000ApJ...529..338R}
{Richer} J.,  {Michaud} G.,   {Turcotte} S.,  2000, \mn@doi [\apj]
  {10.1086/308274}, \href
  {https://ui.adsabs.harvard.edu/abs/2000ApJ...529..338R} {529, 338}

\bibitem[\protect\citeauthoryear{{Ricker} et~al.,}{{Ricker}
  et~al.}{2014}]{1.JATIS.1.1.014003}
{Ricker} G.~R.,  et~al., 2014, \mn@doi [JATIS] {10.1117/1.JATIS.1.1.014003}, 1,
  14003

\bibitem[\protect\citeauthoryear{{Romanovskaya}, {Ryabchikova}, {Shulyak},
  {Perraut}, {Valyavin}, {Burlakova}  \& {Galazutdinov}}{{Romanovskaya}
  et~al.}{2019}]{2019MNRAS.488.2343R}
{Romanovskaya} A.,  {Ryabchikova} T.,  {Shulyak} D.,  {Perraut} K.,  {Valyavin}
  G.,  {Burlakova} T.,   {Galazutdinov} G.,  2019, \mn@doi [\mnras]
  {10.1093/mnras/stz1858}, \href
  {https://ui.adsabs.harvard.edu/abs/2019MNRAS.488.2343R} {488, 2343}

\bibitem[\protect\citeauthoryear{{Romanyuk}}{{Romanyuk}}{2007}]{2007AstBu..62...62R}
{Romanyuk} I.~I.,  2007, \mn@doi [Astrophys. Bull.]
  {10.1134/S1990341307010063}, \href
  {https://ui.adsabs.harvard.edu/abs/2007AstBu..62...62R} {62, 62}

\bibitem[\protect\citeauthoryear{Rucinski, Carroll, Kuschnig, Matthews  \&
  Stibrany}{Rucinski et~al.}{2003}]{RUCINSKI2003371}
Rucinski S.,  Carroll K.,  Kuschnig R.,  Matthews J.,   Stibrany P.,  2003,
  \mn@doi [Adv. Space Res.] {https://doi.org/10.1016/S0273-1177(02)00628-2},
  31, 371

\bibitem[\protect\citeauthoryear{Ryabchikova, Piskunov, Kurucz, Stempels,
  Heiter, Pakhomov  \& Barklem}{Ryabchikova et~al.}{2015}]{Ryabchikova_2015}
Ryabchikova T.,  Piskunov N.,  Kurucz R.~L.,  Stempels H.~C.,  Heiter U.,
  Pakhomov Y.,   Barklem P.~S.,  2015, \mn@doi [Phys. Scr.]
  {10.1088/0031-8949/90/5/054005}, 90, 054005

\bibitem[\protect\citeauthoryear{{Saio}}{{Saio}}{1981}]{1981ApJ...244..299S}
{Saio} H.,  1981, \mn@doi [\apj] {10.1086/158708}, \href
  {https://ui.adsabs.harvard.edu/abs/1981ApJ...244..299S} {244, 299}

\bibitem[\protect\citeauthoryear{{Saio}, {Kurtz}, {Murphy}, {Antoci}  \&
  {Lee}}{{Saio} et~al.}{2018}]{2018MNRAS.474.2774S}
{Saio} H.,  {Kurtz} D.~W.,  {Murphy} S.~J.,  {Antoci} V.~L.,   {Lee} U.,  2018,
  \mn@doi [\mnras] {10.1093/mnras/stx2962}, \href
  {http://adsabs.harvard.edu/abs/2018MNRAS.474.2774S} {474, 2774}

\bibitem[\protect\citeauthoryear{{Santos}, {Israelian}, {Mayor}, {Bento},
  {Almeida}, {Sousa}  \& {Ecuvillon}}{{Santos}
  et~al.}{2005}]{2005A&A...437.1127S}
{Santos} N.~C.,  {Israelian} G.,  {Mayor} M.,  {Bento} J.~P.,  {Almeida} P.~C.,
   {Sousa} S.~G.,   {Ecuvillon} A.,  2005, \mn@doi [\aap]
  {10.1051/0004-6361:20052895}, \href
  {https://ui.adsabs.harvard.edu/abs/2005A&A...437.1127S} {437, 1127}

\bibitem[\protect\citeauthoryear{{Sekiguchi} \& {Fukugita}}{{Sekiguchi} \&
  {Fukugita}}{2000}]{2000AJ....120.1072S}
{Sekiguchi} M.,  {Fukugita} M.,  2000, \mn@doi [AJ] {10.1086/301490}, \href
  {http://adsabs.harvard.edu/abs/2000AJ....120.1072S} {120, 1072}

\bibitem[\protect\citeauthoryear{{Shulyak}, {Tsymbal}, {Ryabchikova},
  {St{\"u}tz}  \& {Weiss}}{{Shulyak} et~al.}{2004}]{2004A&A...428..993S}
{Shulyak} D.,  {Tsymbal} V.,  {Ryabchikova} T.,  {St{\"u}tz} C.,   {Weiss}
  W.~W.,  2004, \mn@doi [\aap] {10.1051/0004-6361:20034169}, \href
  {https://ui.adsabs.harvard.edu/abs/2004A&A...428..993S} {428, 993}

\bibitem[\protect\citeauthoryear{{Skrutskie} et~al.,}{{Skrutskie}
  et~al.}{2006}]{2006AJ....131.1163S}
{Skrutskie} M.~F.,  et~al., 2006, \mn@doi [\aj] {10.1086/498708}, \href
  {http://adsabs.harvard.edu/abs/2006AJ....131.1163S} {131, 1163}

\bibitem[\protect\citeauthoryear{{Smalley}}{{Smalley}}{1993}]{1993A&A...274..391S}
{Smalley} B.,  1993, \aap, \href
  {https://ui.adsabs.harvard.edu/abs/1993A&A...274..391S} {274, 391}

\bibitem[\protect\citeauthoryear{{Smith}}{{Smith}}{1997}]{1997A&A...319..928S}
{Smith} K.~C.,  1997, \aap, \href
  {https://ui.adsabs.harvard.edu/abs/1997A&A...319..928S} {319, 928}

\bibitem[\protect\citeauthoryear{{Sneden}, {Bean}, {Ivans}, {Lucatello}  \&
  {Sobeck}}{{Sneden} et~al.}{2012}]{2012ascl.soft02009S}
{Sneden} C.,  {Bean} J.,  {Ivans} I.,  {Lucatello} S.,   {Sobeck} J.,  2012,
  {MOOG: LTE line analysis and spectrum synthesis}, Astrophysics Source Code
  Library (\mn@eprint {ascl} {1202.009})

\bibitem[\protect\citeauthoryear{{Soufi}, {Goupil}  \& {Dziembowski}}{{Soufi}
  et~al.}{1998}]{1998A&A...334..911S}
{Soufi} F.,  {Goupil} M.~J.,   {Dziembowski} W.~A.,  1998, \aap, \href
  {https://ui.adsabs.harvard.edu/abs/1998A%26A...334..911S} {334, 911}

\bibitem[\protect\citeauthoryear{{Sousa} et~al.,}{{Sousa}
  et~al.}{2015}]{2015A&A...576A..94S}
{Sousa} S.~G.,  et~al., 2015, \mn@doi [\aap] {10.1051/0004-6361/201425227},
  \href {https://ui.adsabs.harvard.edu/abs/2015A&A...576A..94S} {576, A94}

\bibitem[\protect\citeauthoryear{{Stateva}, {Iliev}, {Budaj}  \&
  {Barzova}}{{Stateva} et~al.}{2009}]{stateva}
{Stateva} I.~K.,  {Iliev} I.~K.,  {Budaj} J.,   {Barzova} I.~S.,  2009, BAJ,
  \href {http://adsabs.harvard.edu/abs/2009BlgAJ..12...29S} {12, 29}

\bibitem[\protect\citeauthoryear{{Takeda}, {Han}, {Kang}, {Lee}  \&
  {Kim}}{{Takeda} et~al.}{2008}]{2008JKAS...41...83T}
{Takeda} Y.,  {Han} I.,  {Kang} D.-I.,  {Lee} B.-C.,   {Kim} K.-M.,  2008,
  \mn@doi [J. Korean Astron.] {10.5303/JKAS.2008.41.4.083}, \href
  {https://ui.adsabs.harvard.edu/abs/2008JKAS...41...83T} {41, 83}

\bibitem[\protect\citeauthoryear{{Takeda}, {Kang}, {Han}, {Lee}  \&
  {Kim}}{{Takeda} et~al.}{2009}]{2009PASJ...61.1165T}
{Takeda} Y.,  {Kang} D.-I.,  {Han} I.,  {Lee} B.-C.,   {Kim} K.-M.,  2009,
  \mn@doi [\pasj] {10.1093/pasj/61.5.1165}, \href
  {https://ui.adsabs.harvard.edu/abs/2009PASJ...61.1165T} {61, 1165}

\bibitem[\protect\citeauthoryear{{Talon}, {Richard}  \& {Michaud}}{{Talon}
  et~al.}{2006}]{2006ApJ...645..634T}
{Talon} S.,  {Richard} O.,   {Michaud} G.,  2006, \mn@doi [ApJ]
  {10.1086/504066}, \href
  {https://ui.adsabs.harvard.edu/abs/2006ApJ...645..634T} {645, 634}

\bibitem[\protect\citeauthoryear{{Th{\'e}ado}, {Vauclair}, {Alecian}  \&
  {LeBlanc}}{{Th{\'e}ado} et~al.}{2011}]{2011sf2a.conf..253T}
{Th{\'e}ado} S.,  {Vauclair} S.,  {Alecian} G.,   {LeBlanc} F.,  2011, in
  {Alecian} G.,  {Belkacem} K.,  {Samadi} R.,   {Valls-Gabaud} D.,  eds,
  SF2A-2011: Proceedings of the Annual meeting of the French Society of
  Astronomy and Astrophysics. pp 253--256

\bibitem[\protect\citeauthoryear{{Tkachenko}, {Van Reeth}, {Tsymbal}, {Aerts},
  {Kochukhov}  \& {Debosscher}}{{Tkachenko} et~al.}{2013}]{2013A&A...560A..37T}
{Tkachenko} A.,  {Van Reeth} T.,  {Tsymbal} V.,  {Aerts} C.,  {Kochukhov} O.,
  {Debosscher} J.,  2013, \mn@doi [\aap] {10.1051/0004-6361/201322532}, \href
  {https://ui.adsabs.harvard.edu/abs/2013A&A...560A..37T} {560, A37}

\bibitem[\protect\citeauthoryear{Torres}{Torres}{2010}]{Torres_2010}
Torres G.,  2010, \mn@doi [\aj] {10.1088/0004-6256/140/5/1158}, 140, 1158

\bibitem[\protect\citeauthoryear{{Trust}, {Jurua}, {De Cat}  \&
  {Joshi}}{{Trust} et~al.}{2020}]{2020MNRAS.492.3143T}
{Trust} O.,  {Jurua} E.,  {De Cat} P.,   {Joshi} S.,  2020, \mn@doi [\mnras]
  {10.1093/mnras/stz3623}, \href
  {https://ui.adsabs.harvard.edu/abs/2020MNRAS.492.3143T} {492, 3143}

\bibitem[\protect\citeauthoryear{{Tsymbal}, {Ryabchikova}  \&
  {Sitnova}}{{Tsymbal} et~al.}{2019}]{2019ASPC..518..247T}
{Tsymbal} V.,  {Ryabchikova} T.,   {Sitnova} T.,  2019, in {Kudryavtsev} D.~O.,
   {Romanyuk} I.~I.,   {Yakunin} I.~A.,  eds,  Astronomical Society of the
  Pacific Conference Series Vol. 518, Physics of Magnetic Stars. pp 247--252

\bibitem[\protect\citeauthoryear{{Turcotte}}{{Turcotte}}{2002}]{2002ASPC..259..258T}
{Turcotte} S.,  2002, in {Aerts} C.,  {Bedding} T.~R.,
  {Christensen-Dalsgaard} J.,  eds,  Astronomical Society of the Pacific
  Conference Series Vol. 259, IAU Colloq. 185: Radial and Nonradial Pulsationsn
  as Probes of Stellar Physics. p.~258 (\mn@eprint {} {astro-ph/0111179})

\bibitem[\protect\citeauthoryear{{Turcotte}}{{Turcotte}}{2003}]{2003ASPC..305..199T}
{Turcotte} S.,  2003, in {Balona} L.~A.,  {Henrichs} H.~F.,   {Medupe} R.,
  eds,  Astronomical Society of the Pacific Conference Series Vol. 305,
  Magnetic Fields in O, B and A Stars: Origin and Connection to Pulsation,
  Rotation and Mass Loss. p.~199 (\mn@eprint {} {astro-ph/0304424})

\bibitem[\protect\citeauthoryear{{Vick}, {Michaud}, {Richer}  \&
  {Richard}}{{Vick} et~al.}{2010}]{2010A&A...521A..62V}
{Vick} M.,  {Michaud} G.,  {Richer} J.,   {Richard} O.,  2010, \mn@doi [\aap]
  {10.1051/0004-6361/201014307}, \href
  {https://ui.adsabs.harvard.edu/abs/2010A&A...521A..62V} {521, A62}

\bibitem[\protect\citeauthoryear{{Vick}, {Michaud}, {Richer}  \&
  {Richard}}{{Vick} et~al.}{2011}]{2011A&A...526A..37V}
{Vick} M.,  {Michaud} G.,  {Richer} J.,   {Richard} O.,  2011, \mn@doi [\aap]
  {10.1051/0004-6361/201015533}, \href
  {https://ui.adsabs.harvard.edu/abs/2011A&A...526A..37V} {526, A37}

\bibitem[\protect\citeauthoryear{{Watson}}{{Watson}}{1970}]{1970ApJ...162L..45W}
{Watson} W.~D.,  1970, \mn@doi [ApJL] {10.1086/180620}, \href
  {http://adsabs.harvard.edu/abs/1970ApJ...162L..45W} {162, L45}

\bibitem[\protect\citeauthoryear{{Wenger} et~al.,}{{Wenger}
  et~al.}{2000}]{2000A&AS..143....9W}
{Wenger} M.,  et~al., 2000, \mn@doi [A\&AS] {10.1051/aas:2000332}, \href
  {http://adsabs.harvard.edu/abs/2000A%26AS..143....9W} {143, 9}

\bibitem[\protect\citeauthoryear{{Wright} et~al.,}{{Wright}
  et~al.}{2010}]{2010AJ....140.1868W}
{Wright} E.~L.,  et~al., 2010, \mn@doi [\aj] {10.1088/0004-6256/140/6/1868},
  \href {https://ui.adsabs.harvard.edu/abs/2010AJ....140.1868W} {140, 1868}

\bibitem[\protect\citeauthoryear{{Zhao}, {Zhao}, {Chu}, {Jing}  \&
  {Deng}}{{Zhao} et~al.}{2012}]{2012RAA....12..723Z}
{Zhao} G.,  {Zhao} Y.-H.,  {Chu} Y.-Q.,  {Jing} Y.-P.,   {Deng} L.-C.,  2012,
  \mn@doi [Research in Astronomy and Astrophysics]
  {10.1088/1674-4527/12/7/002}, \href
  {https://ui.adsabs.harvard.edu/abs/2012RAA....12..723Z} {12, 723}

\makeatother
\end{thebibliography}
%
%
%
%

%
%
%
%
%

 \bsp	
 \label{lastpage}
 \end{document}